\documentclass[lettersize,journal]{IEEEtran}

\usepackage{cite}
\usepackage{amsmath,amssymb,amsfonts}
\usepackage{graphicx,subfigure,epstopdf}
\usepackage{textcomp}
\usepackage[colorlinks,linkcolor=black]{hyperref}
\usepackage{algorithm}
\usepackage{algorithmicx}
\usepackage{algpseudocode}  
\usepackage{amsmath}  
\usepackage{multirow}
\usepackage[T1]{fontenc} 

\usepackage{amsthm} 

\newtheorem{property}{Property}

\usepackage[dvipsnames]{xcolor}

\def\BibTeX{{\rm B\kern-.05em{\sc i\kern-.025em b}\kern-.08em
		T\kern-.1667em\lower.7ex\hbox{E}\kern-.125emX}}

\begin{document}
\title{Integrated Sensing, Computing, Communication, and Control for Time-Sequence-Based Semantic Communications}

\author{{ Qingliang Li,~\IEEEmembership{Graduate Student Member,~IEEE,}
		Bo Chang,~\IEEEmembership{Member,~IEEE,} Weidong Mei,~\IEEEmembership{Member,~IEEE,} and Zhi Chen,~\IEEEmembership{Senior~Member,~IEEE}
	\vspace{-1cm}
	}

\thanks{The authors are with the National Key Laboratory of Wireless Communications, University of Electronic Science and Technology of China (UESTC), Chengdu, 611731, China. (e-mail: liqingliang@std.uestc.edu.cn; changb3212@163.com; wmei@uestc.edu.cn; chenzhi@uestc.edu.cn)}

}

\maketitle
\thispagestyle{empty}
\begin{abstract}
In the upcoming industrial internet of things (IIoT) era, a surge of task-oriented applications will rely on real-time wireless control systems (WCSs). For these systems, ultra-reliable and low-latency wireless communication will be crucial to ensure the timely transmission of control information. 
To achieve this purpose, we propose a novel time-sequence-based semantic communication paradigm, where an integrated sensing, computing, communication, and control (ISC3) architecture is developed to make sensible semantic inference (SI) for the control information over time sequences, enabling adaptive control of the robot. 
However, due to the causal correlations in the time sequence, the control information does not present the Markov property. 
To address this challenge, we compute the mutual information of the control information sensed at the transmitter (Tx) over different time and identify their temporal semantic correlation via a semantic feature extractor (SFE) module. By this means, highly correlated information transmission can be avoided, thus greatly reducing the communication overhead. Meanwhile, a semantic feature reconstructor (SFR) module is employed at the receiver (Rx) to reconstruct the control information based on the previously received one if the information transmission is not activated at the Tx. Furthermore, a control gain policy is also employed at the Rx to adaptively adjust the control gain for the controlled target based on several practical aspects such as the quality of the information transmission from the Tx to the Rx. 
We design the neural network structures of the above modules/policies and train their parameters by a novel hybrid reward multi-agent deep reinforcement learning framework. 
On-site experiments are conducted to evaluate the performance of our proposed method in practice, which shows significant gains over other baseline schemes.

\end{abstract}


\section{Introduction}
In the forthcoming era of industrial internet of things (IIoT), a plethora of task-oriented applications are poised to emerge, as propelled by real-time wireless control systems (WCSs) \cite{iiot2018,iiot2020}, e.g., robotic teleoperation and autonomous driving. In such systems, achieving good control performance necessitates ultra-reliable and low-latency communications (URLLC) to guarantee the timely transmission of control information, which, however, pose significant challenges for conventional wireless communication systems, which predominately focus on the accuracy of information bit transmission without accounting for the priority behind the information bits  \cite{ma2019high,wireless_control,wireless_control_commu}. As a result, all generated information bits are treated equally and transmitted with the same effort, leading to exorbitant energy consumption for information transmission and even congestion of the network \cite{IB_Shao}. Notably, various data-hungry applications, e.g., 8K ultra high definition (UHD) video transmission in augmented reality (AR), virtual reality (VR), metaverse, swarm robotics, etc. \cite{SC_2021CM,kountouris2021semantics,zhang2022toward,SC2023} produce substantial data amount, thus aggravating the above network congestion issue, which is known as {imperfect communication}.
These imperfect communication inevitably endanger the normal operations of WCSs.

	\begin{figure*}[!t]
	\centering
	\vspace{-0.cm}
	\subfigure[Traditional WCS.]{
		\setlength{\abovecaptionskip}{-0.1cm}
		\setlength{\belowcaptionskip}{-0.1cm} 
		\includegraphics[scale=0.65]{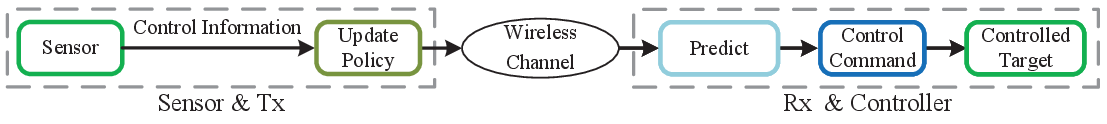}
		\label{fig_tra}
	}
	\subfigure[Time-sequence-based SC-enabled WCS.]{
		\setlength{\abovecaptionskip}{-0.1cm}
		\setlength{\belowcaptionskip}{-0.1cm} 
		\includegraphics[scale=0.65]{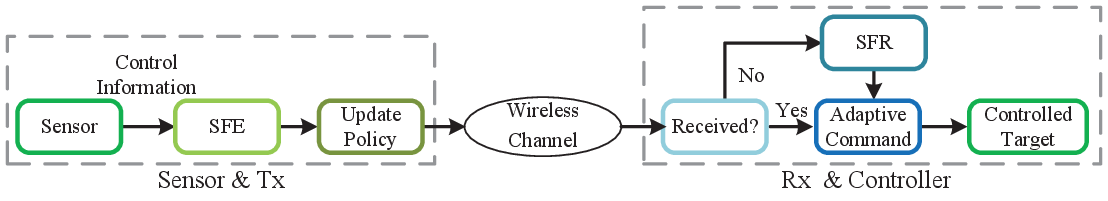}
		\label{fig_toc}
	}
	\caption {Comparison between the traditional WCS and the proposed time-sequence-based SC-enabled WCS.}\label{fig_introduction}
	\vspace{-0.6cm}
\end{figure*}

To tackle the imperfect communication issues in WCSs, a widely adopted method is by identifying the temporal features of the control information for more efficient transmission to reduce the communication overhead. As shown in Fig. \ref{fig_tra}, an information update policy can be applied at the transmitter (Tx) to dynamically adjust the frequency of information transmission based on different criteria. Moreover, a prediction module is applied at the receiver (Rx) to predict the control information in a certain period, if the control information is less frequently transmitted by the Tx. Under this architecture, some existing works have delved into the prediction module design at the Rx and/or the information update policy at the Tx.
For instance, in \cite{hou2019prediction}, a linear prediction model was utilized at the Rx to predict the missing control information.
Furthermore, a combined design of sampling and prediction was proposed in \cite{meng_zhen} to develop an update strategy for metaverse.
Moreover, a task-oriented cross-system design framework integrating sensing, communication, prediction, control, and rendering was established to model a robotic arm in the metaverse in \cite{mengzhen_jsac24}.
In \cite{tongxin}, the increment of information (IoI) was adopted as a criterion to determine the information update policy at the Tx, where only the information with sufficiently high IoI will be transmitted.

Although the above methods are generally effective in reducing the communication overhead, they may not achieve optimal performance since all control information is still treated equally, without identifying their semantic meaning, which, however, may be further leveraged for performance improvement. 
Recently, semantic communications (SCs) \cite{rahman2020comprehensive,lan2021semantic,liu2023survey} have been proposed as a promising solution to identify important features of various types of information. 
In particular, SC extracts semantic information (of low data size) from the bit information (of large data size) by invoking machine learning technologies. By this means, the total amount of information transmission can be considerably reduced as compared to the conventional bit transmission. As an additional merit, the SC can be applied to any data type one intends to communicate, besides the traditional data types, e.g., text\cite{SC_text}, image\cite{SC_image}, and audio\cite{SC_audio}, known as {\it data-oriented SC}. However, the data-oriented SC needs to be redesigned in WCSs, as the control information is generally concise enough already, which may not be further semantically compressed. 
Instead, it is more critical to extract their important temporal features over time, termed {\it time-sequence-based SC}, which belongs to the {\it task-oriented SC}.
In \cite{TCOM24-WCS}, task-oriented SC and rate splitting techniques were used to improve the transmission efficiency and URLLC performance in WCSs.
The authors in \cite{girgis2024semantic} employed dynamic semantic Koopman and logical semantic Koopman to extract the semantic features of the control system and encode control instructions with different rules to reduce the communication overhead in multi-objective control.
Although there are a few prior works on the joint design of semantic communication and wireless control systems, an efficient time-sequence-based SC paradigm has yet to be established.
Unlike existing SC schemes that extract semantic features of control information within a single time slot, our time-sequence-based SC paradigm emphasizes the extraction of causal semantic features across the temporal dimension.


Motivated by the above, we propose in this paper a new time-sequence-based SC paradigm for WCSs by integrating different computing modules at the Tx and the Rx, as shown in Fig. \ref{fig_toc}, which aim to make sensible semantic inference (SI) for the control information over time, such that the controlled target can take a right action at the right time and in the right context.
As compared to the traditional WCS shown in Fig. \ref{fig_tra}, the temporal features of control information can be more effectively identified and leveraged. 
To the best of our knowledge, this paper is the first to explore time-sequence-based SC paradigm using the integrated sensing, computing, communication, and control (ISC3) architecture for the WCS, which helps reduce communication overhead and improve control accuracy. 
Furthermore, the implementation of our proposed scheme relies on established communication systems and does not necessitate the additional design of semantic encoding and modulation techniques.
This approach could provide a feasible paradigm for future SCs in task-oriented applications.

  The main contributions of this paper are summarized as follows.
  \begin{itemize}
   \item{We propose a time-sequence-based SC paradigm by formulating a novel ISC3 architecture in task-oriented applications, as shown in Fig. \ref{fig_toc}. 
   	In particular, at the Tx, we employ an artificial intelligence (AI)-empowered semantic feature extractor (SFE) module to infer the correlation between control information over time and thereby determine its update policy. 
   	While at the Rx, an AI-empowered semantic feature reconstructor (SFR) module is employed to predict the control information at the current time based on the previously received one if it is not transmitted.
   	Furthermore, an AI-empowered control gain policy is also employed at the Rx to adaptively adjust the control gain for the controlled target based on various practical factors, such as the quality of the control information transmission from the Tx to the Rx and the accuracy of the control information prediction at the Rx.
   	Unlike URLLC, which improves wireless control performance by increasing communication resources to minimize latency and enhance reliability, our proposed scheme ensures wireless control performance with minimal resource consumption.
   }
   
   	\item{
   	However, obtaining the optimal transmitting decision by the sensed control state directly is extremely challenging due to its failure to adhere to the Markov property. Specifically, the decision is affected not only by the current state but also by previous states. To address this challenge, we integrate mutual information (MI) and long short-term memory (LSTM) networks with multi-agent deep reinforcement learning (MADRL). 
   	To characterize the wireless control performance, we propose a novel reward function that effectively integrates communication cost and control accuracy.
   	Accordingly, we aim to maximize this reward by jointly optimizing the SFE module, the SFR module, and the control gain policy. 
   	To tackle this problem, we design the neural network structures of these modules and policies and train their parameters by a novel hybrid reward MADRL (HR-MADRL) framework characterized by a mixed action space and joint rewards.
   	 }
   	
   	\item{Finally, we conduct on-site experiments to evaluate the performance of our proposed method and other baseline methods in practice. The experimental results demonstrate that our proposed method can significantly outperform the baseline methods and achieve a much better trade-off between the communication overhead and the control performance.
  }
\end{itemize}

The rest of this paper is organized as follows. In Section II, we present the components of the ISC3 system and their system models, as well as the problem formulation. In Section III, we present the proposed AI-based solution to the formulated problem. In Section IV, we present our experimental results. Section V concludes the paper.

The following notations are utilized throughout this paper.
Bold symbols in lowercase and uppercase denote vectors and matrices, respectively.
$\mathbb{R}^{n \times m}$ denotes the set of all $n \times m$ real matrices. 
$\boldsymbol{I}_{n}$ denotes an $n \times n$ identity matrix. 
$\mathbb{P}_{XZ}$ denotes the joint probability distribution of random variables $X$ and $Z$. $\mathbb{P}_{X} \otimes \mathbb{P}_{Z}$ denotes the product of the marginals probability distribution. 
$X \sim {\cal N}(0,\sigma^2)$ means that the random variables $X$ follows a zero-mean Gaussian distribution with variance $\sigma^2$. 
$X \sim B(p)$ means that $X$ follows Bernoulli distribution with parameter $p$.
$\Vert \boldsymbol{x} \Vert$ denotes the Euclidean norm of the vector $\boldsymbol{x}$. $\mathbb{E}[X]$ denotes the expectation of a random variable $X$. $\Delta_{\boldsymbol{x}} y$ denotes the gradient of $y$ with respect to $\boldsymbol{x}$. $a \leftarrow b$ denotes that the value of $b$ is assigned to $a$. 
``$\rm mod$'' denotes the remainder operator. 
To facilitate quick reference, the main symbols and their descriptions used throughout this paper are presented in Table \ref{symbol}.


\begingroup
\allowdisplaybreaks
\section{System Model and Problem Formulation}

In this section, we provide system model for the proposed ISC3 scheme to achieve the time-sequence-based SC architecture. 
We focus on a time-sequence-based and SC-enabled WCS, as shown in Fig. \ref{fig_toc}. The Tx updates its decisions by analyzing the correlation between current and previously sent control information. Notably, this policy relies on both immediate and historical states, which violates the Markov property. In particular, this non-Markovian nature requires additional contextual information, rendering the decision-making more complicated. On the other hand, the Rx needs to infer complete control command sequences based on its received sparse control information. This is similar to a generation problem with indefinite length and may be addressed using generative models such as the Transformer. However, for real-time control systems, deploying large models like Generative Pre-trained Transformer (GPT) are impractical due to high computational costs. 
Therefore, lightweight computational models are more desired to balance resource efficiency and performance. 
This study proposes measures to enhance the system's Markovian properties and simplify the computational structure to meet control task constraints.
Next, we first present the components of our proposed WCS with ISC3, followed by their respective system models and the overall problem formulation.

\begin{figure*}[!t]
	\centering
	\vspace{-0.cm}
	\setlength{\abovecaptionskip}{-0.0cm}
	\setlength{\belowcaptionskip}{-0.0cm} 
	\includegraphics[scale=0.75]{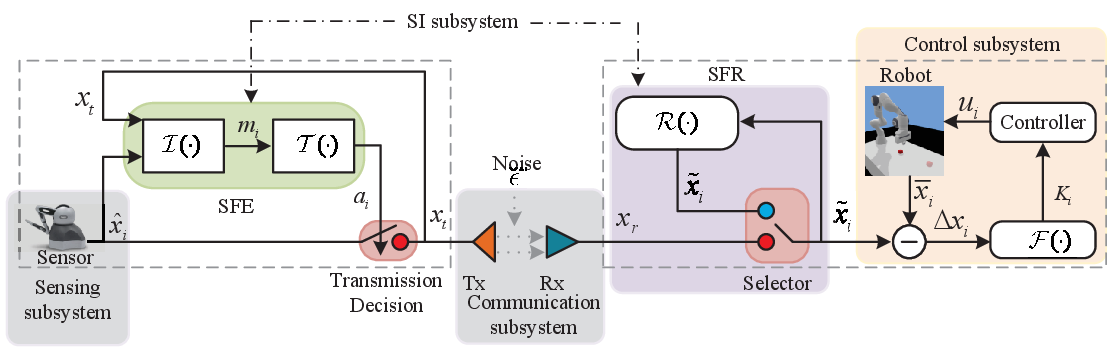}
	\caption{Proposed WCS with ISC3.}
	\vspace{-0.2cm}
	\label{system_diagram}
\end{figure*}
	
	
\subsection{Overall Architecture}

As shown in Fig. \ref{system_diagram}, our proposed architecture for WCS consists of four subsystems corresponding to sensing, communication, control, and SI, respectively. The SI subsystem is equipped with an SFE computing module and an SFR computing module at the Tx and Rx, respectively. The workflow of our proposed ISC3 system is delineated as follows.
First, a Geomagic Touch X device is deployed as a sensor in the sensing subsystem, sending its sensed  control information to the SFE computing module. After AI inference, a decision is made to determine the information update policy, i.e., whether the control information should be transmitted to the Rx at this moment.
Particularly, if it is highly correlated with the previously transmitted information, it may not be transmitted to reduce the communication overhead. Otherwise, it can be transmitted. 
In the latter case, the transmitted information will directly control the target device, i.e., a Franka Emika Robot Arm. While in the former case without transmission, the SI subsystem will predict and reconstruct the control information based on the previously received one. In the following, we elaborate upon the four subsystems.

\begin{table}
	\caption{Symbols and Descriptions}
	\centering
	\begin{tabular}{|c|l|}
		\hline
		\textbf{Symbol} & \textbf{Description} \\
		\hline
		$a_{i}$ & Transmission decision at time slot $i$\\
		$e_{i}$ & Control error at time slot $i$\\
		$m_{i}$ & MI between $\boldsymbol{\hat x}_i$ and $\boldsymbol{x}_t$  \\
		$u_{i}$ & Control command at time slot $i$ \\
		$v_i$ & Huber Loss at time slot $i$ \\
		$r_i$ & Reward of update policy at time slot $i$\\
		$\boldsymbol{x}_i$ & Desired control information at time slot $i$\\
		$\boldsymbol{\hat x}_i$ & Sensed control information at time slot $i$ \\
		$\boldsymbol{x}_t$ & Last information sent to Rx\\
		$\boldsymbol{x}_r$ & Last information received at Rx  \\
		$\boldsymbol{\tilde{x}}_i$ & Decoded control information at time slot $i$\\
		$\boldsymbol{\bar{x}}_i$ & Status of the robot at time slot $i$ \\
		$K_i$ & Robotic control gain at time slot $i$  \\
		$\mathcal{F}(\cdot)$ & Robotic control gain policy \\
		$\mathcal{G}(\cdot)$ & Robot arm drive system  \\
		$\mathcal{I}(\cdot)$ & Semantic feature extractor (SFE)\\
		$\mathcal{R}(\cdot)$ & Semantic feature reconstructor (SFR)\\
		$\mathcal{T}(\cdot)$ & Semantic information update policy\\
		$\delta_l$ & Control accuracy parameter\\
		$\delta_u$ & Control-error upper bound\\
		$\epsilon$ & Status noise \\
		\hline
	\end{tabular}\label{symbol}
\end{table}

\subsection{Sensing Subsystem}

As shown in Fig. \ref{system_diagram}, with the Geomagic Touch X device deployed as a sensor, the operator utilizes its Touch pen to execute 3D drawings, whereby the real-time positional data of the Touch nib is sampled and utilized as control input for wirelessly controlling the robotic arm. This prototype can find potential applications in various practical scenarios, such as telemedicine and the metaverse.
Let the desired control information at the Tx in any time slot $i$ be denoted as $\boldsymbol{x}_i \in \mathbb{R}^{n_x \times 1}$, where $n_x$ is the total number of states. Then, the actual control information sensed by the sensor is expressed as 
\begin{equation} \label{equ1}
	\boldsymbol{\hat x}_i=\boldsymbol{x}_i+\boldsymbol{\epsilon}_s,
\end{equation}
where $\boldsymbol{\epsilon}_s \sim {\cal N}(\boldsymbol{0},\sigma_s^2 \boldsymbol{I}_{n_x})$ is the additive white Gaussian noise (AWGN) over the sensing channel. 
Given that the state under consideration is represented in a three-dimensional coordinate system with consistent units of measurement and similar orders of magnitude across all dimensions, we simplify the modeling process by assuming that all states noise are independent and identically distributed random variables with identical variance.
In addition, since the data collected by the sensors are inherently high-precision and structurally concise, we do not allocate additional computing resources to compress each frame of sensing data. Instead, we implement a temporal sampling mechanism to extract data that is rich in semantic information for transmission. Relevant details will be elaborated upon in Sections II.C and II.D.

\subsection{Communication Subsystem}
To maintain generality, we consider a generic communication scenario. In this scenario, not all data sampled by the sensors are transmitted; instead, redundant data are discarded to reduce the communication load. It is important to note that this triggering decision mechanism is applicable to a variety of communication scenarios, including wired communication, wireless communication, and network communication systems that utilize different protocols.
Let $a_i \in \{0,1\}$ denote the transmission decision at time slot $i$. If $a_i=1$, the Tx will send the control information to the Rx. Otherwise, the transmission will not be activated. Accordingly, we define the duty cycle (DC) of the wireless channel use as 
\begin{equation} \label{equ2}
	\eta_{dc}=\frac{\mathbb{I} (a_i=1) }{\mathbb{I} (a_i=1)+\mathbb{I} (a_i=0)},
\end{equation}
where $\mathbb{I} (a_i=1) $ means the total number of activations of the information transmission or channel use, and $\mathbb{I} (a_i=0) $ means that of inactivations of the channel use. 
In the case with $a_i=1$, the received information at the Rx is denoted as 
\begin{equation}
	\boldsymbol{x}_r =\boldsymbol{\hat{x}}_i+\boldsymbol{\epsilon}_c = \boldsymbol{x}_i + \boldsymbol{\epsilon}_s + \boldsymbol{\epsilon}_c,
	\label{equ6}
\end{equation}
where $\boldsymbol{\epsilon}_c \sim {\cal N}(\boldsymbol{0},\sigma_c^2 \boldsymbol{I}_{n_x})$ is the AWGN at the receiver.
Accordingly, we define the overall AWGN as 
\begin{equation}
	\boldsymbol{\epsilon} =\boldsymbol{\epsilon}_s + \boldsymbol{\epsilon}_c= \boldsymbol{x}_r - \boldsymbol{x}_i,
	\label{equ7}
\end{equation}
with $\boldsymbol{\epsilon} \sim {\cal N}(\boldsymbol{0},\sigma^2 \boldsymbol{I}_{n_x})$.

\subsection{SI Subsystem}
As shown in Fig. \ref{system_diagram}, the SI subsystem consists of an SFE computing module at the Tx to make the transmission decision and an SFR computing module at the Rx to reconstruct the information needed in the control subsystem. 

For the SFE at the Tx, let the last sensed control information that has been transmitted to the Rx be expressed as $\boldsymbol{x}_t$, with $t < i$. To characterize the correlation between $\boldsymbol{\hat x}_i$ and $\boldsymbol{x}_t$, we propose to adopt their MI, which is expressed as\cite{emi}
\begin{align}
	&I(\boldsymbol{\hat x}_i, \boldsymbol{x}_t)  = H(\boldsymbol{\hat x}_i) - H(\boldsymbol{\hat x}_i \vert \boldsymbol{x}_t) \nonumber\\ 
	&=
		\!\!-\!\! \int_{\boldsymbol{\hat x}_i} \!\!  p(\boldsymbol{\hat x}_i)  
	\log p(\boldsymbol{\hat x}_i) \mathrm{d}\boldsymbol{\hat x}_i
	\!\!\!+\!\!
	\int_{\boldsymbol{\hat x}_i, \boldsymbol{x}_t} \!\!\!\!\!\! 
	p(\boldsymbol{\hat x}_i , \boldsymbol{x}_t)  \log p(\boldsymbol{\hat x}_i \vert \boldsymbol{x}_t)
	\mathrm{d}\boldsymbol{\hat x}_i \mathrm{d}\boldsymbol{x}_t  
\nonumber \\ 
	&= D_{KL} \left(\mathbb{P}_{\boldsymbol{\hat x}_i, \boldsymbol{x}_t} \Vert \mathbb{P}_{\boldsymbol{\hat x}_i} \otimes \mathbb{P}_{\boldsymbol{x}_t}\right),
	\label{equ3}
\end{align}
where $H(\cdot)$ is the Shannon entropy, $D_{KL}(\cdot)$ is the Kullback-Leibler (KL) divergence\cite{dkl}. 
However, computing \eqref{equ3} requires {\it a priori} knowledge about the probability density function $\mathbb{P}_{\boldsymbol{\hat x}_i, \boldsymbol{x}_t}$ and $\mathbb{P}_{\boldsymbol{\hat x}_i} \otimes \mathbb{P}_{\boldsymbol{x}_t}$, which may not be practically available. 
Thus, we adopt a deep neural network (DNN), i.e., Mutual Information Neural Estimator (MINE) \cite{MINE}, denoted as $\mathcal{I}(\cdot)$, to approximate the actual MI. Accordingly, the output value of the MINE at the SFE in time slot $i$ is denoted as
\begin{equation} \label{mi}
	m_i = \mathcal{I}(\boldsymbol{\hat x}_i, \boldsymbol{x}_t),
\end{equation}
where $\mathcal{I}(\boldsymbol{\hat x}_i, \boldsymbol{x}_t)$ is employed to approximate the actual MI $I(\boldsymbol{\hat x}_i, \boldsymbol{x}_t)$.
After the processing by MINE, an update policy $\mathcal{T}(\cdot)$ is then applied to determine whether the control information should be transmitted based on \eqref{mi}, i.e., 
\begin{equation} \label{dqn}
	a_i = \mathcal{T}(m_i)\in \{0,1\}.
\end{equation}
Compared to directly updating strategies based on state $\boldsymbol{\hat x}_i$, using MI as the input state ensures compliance with the Markov decision process (MDP).
In addition, the value of MI can quantify the semantic feature differences between \(\boldsymbol{\hat x}_i\) and \(\boldsymbol{x}_t\), thereby enabling the update policy $\mathcal{T}(\cdot)$ to learn accurate activation time.


The SFR module consists of a feature encoder and a feature decoder to identify the features of the transmitted information by the Tx. In each time slot $i$, if no control information is received from the Tx, i.e., $a_i=0$, the SFR will predict the control information based on that (received or predicted) in time slot $i-1$. If $a_i=1$, it will decode $\boldsymbol{\hat x}_i$ based on \eqref{equ6}. The above process is executed by utilizing the selector shown in Fig. \ref{system_diagram}. Let $\mathcal{R}(\cdot)$ denote the prediction policy. Then, the control information sent to the control subsystem in time slot $i$ is given by
\begin{equation} \label{decoder}
	\boldsymbol{\tilde{x}}_i = \left\{ \begin{array}{ll}
		\mathcal{R} \left(\boldsymbol{\tilde{x}}_{i-1} \right),&{\rm if}\ a_i = 0, \\
		\boldsymbol{x}_r,&{\rm if}\ a_i = 1,
	\end{array} \right.
\end{equation}
where $\boldsymbol{\tilde{x}}_i$ is the decoded control information at time slot $i$. We need to note that the transmission decision at the Tx is mainly determined by the wireless channel condition and the SFE module, without any feedback from the Rx. As such, to achieve \eqref{decoder}, we define a maximum time interval $\tau$ for the Rx. If the Rx does not receive any new information within $\tau$, it will declare that the transmission is not activated, i.e., $a_i = 0$. In this case, the predicted status information by the SFR module is used, i.e., the first case of \eqref{decoder}.


\subsection{Control Subsystem}
As shown in Fig. \ref{system_diagram}, the control subsystem consists of a robotic control gain module $\mathcal{F}(\cdot)$, a controller and a robot arm. 
Based on the status information in \eqref{decoder}, the control command for the robotic arm is given by 
 \begin{equation} \label{controller}
 	\boldsymbol{u}_i = K_i \left(\boldsymbol{\bar{x}}_i - \boldsymbol{\tilde{x}}_i\right) =K_i \Delta \boldsymbol{x}_i ,
 \end{equation}
where $K_i$ and $\boldsymbol{\bar{x}}_i$ denote the robotic control gain and the actual status of the robot arm in time slot $i$, respectively, and $\Delta \boldsymbol{x}_i = \boldsymbol{\bar{x}}_i - \boldsymbol{\tilde{x}}_i$ describes the difference between the actual and desired statues of the robot arm.

Regarding the robotic control gain in \eqref{controller}, it is practically dependent on the status difference $\Delta \boldsymbol{x}_i$, transmission decision $a_i$, and the noise variance $\sigma^2$. Particularly, if the status difference is large, a small control gain is typically sufficient to prevent overshooting of the robot arm drive system, whereas a larger gain is needed if the status difference is small. In addition, the robotic control gain is also affected by the transmission decision and the noise variance. For instance, in the presence of a high noise variance or $a_i=0$, it is preferred to adopt the predicted information by the SFR instead of the received information from the Tx (if $a_i=1$), which helps result in a larger control gain thanks to the more accurate commands. Based on the above, the robotic control gain policy can be expressed as a function, i.e.,
\begin{equation} \label{control_policy}
		K_i = \mathcal{F} \left(\Delta \boldsymbol{x}_i, a_i, \sigma^2 \right),
\end{equation}
where $\mathcal{F}(\cdot)$ captures the effect of $\Delta \boldsymbol{x}_i, a_i$, and $\sigma^2$ on control gain.

Let $\mathcal{G}(\cdot)$ denote the driving policy of the robot arm, which depends on the robot drive system and is fixed. Given the input status information $\boldsymbol{u}_i$, it outputs the actual status in time slot $i+1$, i.e.,
\begin{equation} \label{robot}
	\boldsymbol{\bar{x}}_{i+1} = \mathcal{G} \left( \boldsymbol{u}_i \right).
\end{equation}
  
\subsection{Problem Formulation}
Under the above $\mathrm{ISC}^3$ architecture, we aim to minimize the overall communication overhead while ensuring precise teleoperation of the robot based on the SI design, subject to the constraint on the control error. The associated optimization problem can be formulated as
\begin{equation*}\label{eq:P_1}
	\begin{split}
		{\mathcal{P} }&:  \quad \min_{\mathcal{I},\mathcal{T},\mathcal{R}, \mathcal{F}} \quad \mathbb{E}\left[ \eta_{dc}  \right] \\
		s.t. 
		&\quad  \ \Vert 
			\boldsymbol{\bar{x}}_i - \boldsymbol{\hat x}_i
		\Vert \leq \delta_u, \\
	\end{split}
\end{equation*}
where $\eta_{dc}$ represents the DC of the wireless channel, as defined in \eqref{equ2}; and
	$\Vert 
	\boldsymbol{\bar{x}}_i - \boldsymbol{\hat x}_i
  \Vert$ 
represents the control error between the actual status of the robot arm and the original status from the Geomagic Touch X device in time slot $i$ (see the sensing subsystem in Fig. \ref{system_diagram}), and $\delta_u$ is a prescribed upper bound on the control error.

However, it is difficult to solve $\mathcal{P}$ due to the following reasons. 
First, its optimization needs to calculate the joint MI of multi-dimensional continuous status.
Second, a theoretical model is needed to describe the changes of the robot for SI. Third, it is difficult to optimize the update policy $\mathcal{T}(\cdot)$ and control gain policy $\mathcal{F}(\cdot)$, due to the ``black box'' nature of the input-output relationship of the considered system.
To tackle the above challenges, we propose an AI-based method, as detailed in Section III.

\section{Proposed Solution to $\mathcal{P}$}

The problem $\mathcal{P}$ is a sequential decision-making process with multiple agents. In this process, modules $\mathcal{I}(\cdot)$ and $\mathcal{R}(\cdot)$ provide state information, while modules $\mathcal{T}(\cdot)$ and $\mathcal{F}(\cdot)$ generate decisions. To optimize policies using DRL, we first convert the problem into an MDP. To ensure the system satisfies the Markov property, we train the modules $\mathcal{I}(\cdot)$ and $\mathcal{R}(\cdot)$ using deep learning methods. We then apply MADRL to train agents $\mathcal{T}(\cdot)$ and $\mathcal{F}(\cdot)$. The agent $\mathcal{T}(\cdot)$ has a discrete action space; while agent $\mathcal{F}(\cdot)$ has a continuous action space, and its strategy is developed based on the actions of agent $\mathcal{T}(\cdot)$.
Therefore, in this section, we first design the network structure of $\mathcal{I}(\cdot)$ to calculate the empirical MI. Next, we design the network structures of the prediction policy $\mathcal{R}(\cdot)$. Finally, we employ HR-MADRL to train the associated neural networks.



\subsection{Network Structure of $\mathcal{I}(\cdot)$ for MI Calculation}

Prior to optimizing the semantic information update policy $\mathcal{T}(\cdot)$ at the Tx, it is imperative to design the MINE in the SFE to calculate the empirical MI, i.e., $\mathcal{I}(\cdot)$ in \eqref{mi}.
Specifically, the MINE is a DNN with its parameters denoted as ${\boldsymbol{\theta}}$. For random variables $X$ and $Z$ with $N$ samples, the estimated MI is given by \cite{MINE}
\begin{equation}\label{mi_e}
	\mathcal{V}({\boldsymbol{\theta}}) =\sup _{{\boldsymbol{\theta}}} \mathbb{E}_{\mathbb{P}_{X Z}^{(N)}}\left[\mathcal{I}\right]-\log \left(\mathbb{E}_{\mathbb{P}_X^{(N)} \otimes \mathbb{P}_Z^{(N)}}\left[e^{\mathcal{I}}\right]\right) ,
\end{equation}
where $\mathbb{P}^{(N)}$ represents the empirical distribution associated with $N$ independent identically distributed (i.i.d.) samples. Given a batch of data of size $N$, the estimated gradient of $\mathcal{I}$ can be expressed as
\begin{equation}
	\setlength{\abovedisplayskip}{3pt}
	\setlength{\belowdisplayskip}{3pt}
	\nabla_{\boldsymbol{\theta}} \mathcal{V}({\boldsymbol{\theta}})=\mathbb{E}_N\left[\nabla_{\boldsymbol{\theta}} \mathcal{I}\right]-\frac{\mathbb{E}_N\left[\nabla_{\boldsymbol{\theta}} \mathcal{I} e^{\mathcal{I}}\right]}{\mathbb{E}_N\left[e^{\mathcal{I}}\right]}.
	\label{equ16}
\end{equation}
However, the second term in (\ref{equ16}) will introduce a biased estimate for full batch gradient \cite{MINE}. For example, in an extreme scenario where a batch has only a single sample, we have
\begin{equation}
	\frac{\mathbb{E}_N\left[\nabla_{\boldsymbol{\theta}} \mathcal{I} e^{\mathcal{I}}\right]}{\mathbb{E}_N\left[e^{\mathcal{I}}\right]} \neq \frac{\nabla_{\boldsymbol{\theta}} \mathcal{I} e^{\mathcal{I}}}{e^{\mathcal{I}}} \neq \mathbb{E}_N\left[\nabla_{\boldsymbol{\theta}} \mathcal{I}\right] .
\end{equation}
Thus, it is not feasible to directly employ a batch-based optimization algorithm to optimize this objective. 

To overcome this issue, we leverage the fact that $\mathcal{I}$ does not change drastically in several consecutive iterations. Then, we can adopt a moving average to estimate
$\mathbb{E}_N\left[\nabla_{\boldsymbol{\theta}} \mathcal{I}\right]$. Specifically, we define $\mathbb{X}$ and $\mathbb{X}_r$ as the data sets transmitted at the Tx and received at the Rx, respectively. We randomly sample $N$ batches from these two data sets and denote the $n$-th batch as $\boldsymbol{x}^{(n)}_i$, $\boldsymbol{x}^{(n)}_t \in \mathbb{X}$ and $\boldsymbol{x}^{(n)}_r \in \mathbb{X}_r$, respectively, $n = {1,2,\cdots,N}$. Then, the moving average of the MI in \eqref{mi_e} is given by \cite{MINE}
\begin{equation}\label{mine_loss}
	\setlength{\abovedisplayskip}{3pt}
	\setlength{\belowdisplayskip}{3pt}
	\mathcal{V}({\boldsymbol{\theta}}) \leftarrow \frac{1}{N} \sum_{n=1}^N \mathcal{I}\left(\boldsymbol{x}^{(n)}_i,\boldsymbol{x}_t^{(n)}\right)
	-\frac{\bar{e}}{\bar{e}_0}\log \left(\bar{e}\right),
\end{equation}
where $\bar{e} = \frac{1}{N} \sum_{i=1}^N e^{\mathcal{I}\left(\boldsymbol{x}^{(n)}_i, \boldsymbol{x}_r^{(n)}\right)}$ is the estimated value of $\mathbb{E}_{\mathbb{X}^{(N)} \otimes \mathbb{X}_r^{(N)}}\left[e^{\mathcal{I}}\right]$ and $\bar{e}_0$ is the moving average of $\bar{e}$ and dynamically updated as $\bar{e}_0 \leftarrow \gamma \bar{e}_0 + (1-\gamma) \bar{e}$, with $\gamma \in (0,1)$ denoting a weight parameter. The procedures of our proposed unbiased MINE are summarized in Algorithm \ref{alg:MINE}.



\begin{algorithm}
	\caption {Unbiased MINE Algorithm}
	\label{alg:MINE} 
	\textbf{Initialization}: Initialize the parameters of $\mathcal{I}$, i.e., ${\boldsymbol{\theta}}$. Let $\bar{e}_0 = 1$, $n \in \{1,2,\cdots,N\}$.
	\begin{algorithmic}[1]
		\While{not convergence}
		\State Randomly sample $N$ batches from $\mathbb{X}$: $\boldsymbol{x}^{(n)}_i $;
		\State Randomly sample $N$ batches from $\mathbb{X}$: $\boldsymbol{x}_t^{(n)}$;
		\State Randomly sample $N$ batches from $\mathbb{X}_r$: $\boldsymbol{x}_r^{(n)}$;
		\State Calculate the average: $\bar{e} = \frac{1}{N} \sum_{n=1}^N e^{\mathcal{I}\left(\boldsymbol{x}^{(n)}_i, \boldsymbol{x}_r^{(n)}\right)}$;
		\State Calculate the moving average: $\bar{e}_0 \leftarrow \gamma \bar{e}_0 + (1-\gamma) \bar{e}$;
		\State Calculate the loss $\mathcal{V}({\boldsymbol{\theta}})$ based on \eqref{mine_loss};
		\State Update the bias corrected gradients:  $\nabla {\boldsymbol{\theta}} \leftarrow \nabla_{\boldsymbol{\theta}} \mathcal{V}({\boldsymbol{\theta}})$;
		\State Update the network parameters: ${\boldsymbol{\theta}} \leftarrow {\boldsymbol{\theta}} + \nabla {\boldsymbol{\theta}}$;
		\EndWhile
	\end{algorithmic}  
\end{algorithm}
\vspace{-0.5cm}

\subsection{Network Structure of $\mathcal{R}(\cdot)$}

As shown in (\ref{decoder}), when the control information is not received at the Rx, the predicted status information by the SFR module is used, i.e., $\boldsymbol{\tilde{x}}_i=\mathcal{R} \left(\boldsymbol{\tilde{x}}_{i-1} \right)$. Specifically, the Rx is to reconstruct the semantic features of the motion control information based on the previously received information. In this paper, we apply the LSTM encoder-decoder architecture network to achieve this purpose.
For brevity, the details of the LSTM are omitted, where interested readers can refer to \cite{LSTM}.
In the LSTM network training, we construct an empirical buffer using $N_l$ historical data $\boldsymbol{x}_{r,l},\ l \in [1,N_l]$ consecutively transmitted from the Tx to the Rx.
As shown in Fig. \ref{LSTM}, an encoder reads the input sequence $\boldsymbol{S}_{in} = \left[ \boldsymbol{x}_{r,l+1}, \boldsymbol{x}_{r,l+2}, \cdots, \boldsymbol{x}_{r,l+T}  \right]$ with length $T$ and extracts the semantic feature information to obtain a feature vector. The decoder adopts the output of the encoder as the initial hidden state vector and employs the final input state $\boldsymbol{x}_{r,l+T}$ of the encoder as its input. Finally, the decoder produces a deduced semantic feature information sequence $\boldsymbol{S}_{out} = \left[ \boldsymbol{y}_{1}, \boldsymbol{y}_{2}, \cdots, \boldsymbol{y}_{M} \right]$ with length $M$. In addition, $\boldsymbol{h}_{n_t},\ n_t \in \{0,1,\cdots,T\}$ and $\boldsymbol{h}_j^{\prime},\ j \in \{0,1,\cdots,M-1\}$ are hidden status vectors in Fig. \ref{LSTM}. As such, the  decoder policy $\mathcal{R}(\cdot)$ in \eqref{decoder} selects the first predicted value $\boldsymbol{y}_1$ as the output, which is given by
\begin{equation}
	\mathcal{R}(\boldsymbol{x}_{r,l+T}) = \boldsymbol{y}_1 .
\end{equation}
Although we only utilize the first predicted value in practical applications, we employ a strategy of training the LSTM model to forecast values for the subsequent \( M \) time steps. This approach was implemented to mitigate the “laziness” phenomenon in predictive models, where the model tends to output values that are close to or identical to the last input value, thereby achieving a lower loss value. By adopting this strategy, we encourage the model to learn more predictive features, thereby enhancing its forecasting performance.

It is worth noting that the above semantic feature encoder-decoder characterizes the conditional probability distribution of $\boldsymbol{S}_{in}$ given $\boldsymbol{S}_{out}$. Specifically, given the input sequence $\boldsymbol{S}_{in}$, the conditional probability of the feature vector $\boldsymbol{h}_T$ can be approximated as \cite{LSTM}
\begin{equation}
	p\left(\boldsymbol{h}_T \mid \boldsymbol{S}_{in} \right) \approx  \prod_{n_t=1}^{T} p\left(\boldsymbol{h}_{n_t} \mid \boldsymbol{h}_{{n_t}-1}, \boldsymbol{x}_{r,l+{n_t}} \right) ,
\end{equation}
where the initial hidden vector $\boldsymbol{h}_0$ is randomly generated. As shown in Fig. \ref{LSTM}, the decoder successively produces the probability distribution of the predicted status $\boldsymbol{y}_j$, which is approximated as \cite{LSTM}
\begingroup\makeatletter\def\f@size{9}\check@mathfonts
\def\maketag@@@#1{\hbox{\m@th\normalsize\normalfont#1}}%
\begin{align}
	p\left(\boldsymbol{y}_j \mid \boldsymbol{S}_{in} \right) \approx  \prod_{n_t=1}^{j} p\left(\boldsymbol{h}_{n_t}^{\prime} \mid \boldsymbol{h}_{n_t-1}^{\prime}, \boldsymbol{x}_{r,l+T} \right) 
	  p\left(\boldsymbol{y}_j \mid \boldsymbol{h}_{n_t}^{\prime} \right),
\end{align}
\endgroup
where the initial hidden vector $\boldsymbol{h}_0^{\prime}$ can be obtained from $\boldsymbol{h}_T$ via linear mapping, i.e., $\boldsymbol{h}_0^{\prime} = \boldsymbol{h}_T$.

The performance of the decoder is evaluated by MSE loss, which is given by
\begin{equation}
	L_e(\boldsymbol{S}_{out}, \boldsymbol{S}_{tar}) = \frac{1}{M} \sum_{j=1}^{M} \left(\boldsymbol{y}_j - \boldsymbol{x}_{r,l+T+j}\right)^2,
	\label{equ21}
\end{equation}
where $\boldsymbol{S}_{tar} = \left[\boldsymbol{x}_{r,l+T+1},\boldsymbol{x}_{r,l+T+2},\cdots,\boldsymbol{x}_{r,l+T+M}\right],\ l+T+M\leq N_l$ is the desired output of the decoder.
Then, we can optimize the parameters ${\boldsymbol{\theta}}_{\mathcal{R}}$ in the  LSTM encoder-decoder network to minimize the loss in (\ref{equ21}). The main procedures of the LSTM encoder-decoder algorithm are summarized in Algorithm \ref{alg:LSTM}.

\begin{figure}[!t]
	\centering
	\includegraphics[scale=0.75]{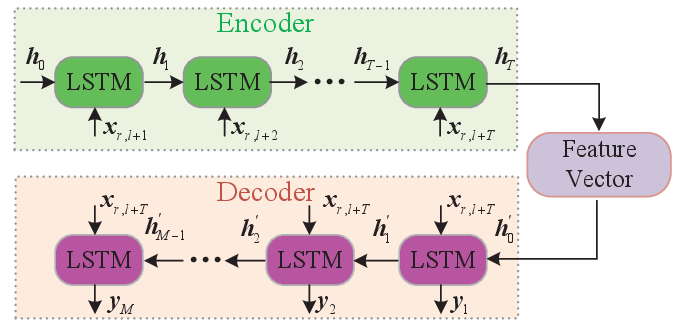}
	\setlength{\abovecaptionskip}{-0.2cm}
	\setlength{\belowcaptionskip}{-0.2cm} 
	\caption{Proposed LSTM encoder-decoder structure.}
	\vspace{-0.1cm}
	\label{LSTM}
\end{figure}

\begin{algorithm}
	\caption {LSTM Encoder-Decoder Algorithm}
	\label{alg:LSTM} 
	\textbf{Initialization}: Initialize the parameters of the LSTM encoder-decoder network, i.e, ${\boldsymbol{\theta}}_{\mathcal{R}}$.
	\begin{algorithmic}[1]
		\While{convergence is not reached}
		\State Randomly sample $N$ batches of length $T+m$ from $\mathbb{X}_r$ as $\boldsymbol{S}^{(n)}=\left[ \boldsymbol{S}_{t}^{(n)},\boldsymbol{S}_{tar}^{(n)} \right], \boldsymbol{S}_{t}^{(n)}\in \mathbb{R}^{n_x\times T}, \boldsymbol{S}_{tar}^{(n)} \in \mathbb{R}^{n_x\times m}, n = {1,2,\cdots,N} $;
		\State Add status noise to the sequence: $\boldsymbol{S}_{in}^{(n)} = \boldsymbol{S}_{t}^{(n)} + \boldsymbol{\epsilon}$;
		\State Input $\boldsymbol{S}_{in}^{(n)}$ into the LSTM encoder-decoder network and obtain its output $\boldsymbol{S}_{out}^{(n)}$;
		\State Calculate the loss: 
		\begingroup\makeatletter\def\f@size{8}\check@mathfonts
		\def\maketag@@@#1{\hbox{\m@th\normalsize\normalfont#1}}%
		$\mathcal{V}({\boldsymbol{\theta}}_{\mathcal{R}}) \leftarrow \frac{1}{N} \sum_{n=1}^N L_e\left(\boldsymbol{S}_{out}^{(n)},\boldsymbol{S}_{tar}^{(n)}\right)$
		\endgroup
		;
		\State Calculate the gradient:  $\nabla {\boldsymbol{\theta}}_{\mathcal{R}} \leftarrow \nabla_{{\boldsymbol{\theta}}_{\mathcal{R}}} \mathcal{V}({\boldsymbol{\theta}}_{\mathcal{R}})$;
		\State Update the network parameters: ${\boldsymbol{\theta}}_{\mathcal{R}} \leftarrow {\boldsymbol{\theta}}_{\mathcal{R}} + \nabla {\boldsymbol{\theta}}_{\mathcal{R}}$;
		\EndWhile
	\end{algorithmic}  
\end{algorithm}
\vspace{-0.4cm}


\subsection{Hybrid Reward Multi-Agent Deep Reinforcement Learning}

To solve the problem $\mathcal{P}$, we model the problem $\mathcal{P}$ as a MDP, which includes the state $S$, action $A$, and reward $R$ of the system.

\subsubsection{State Space}

For policy networks, we expect decisions to be made based on the MI between current and transmitted control information, temporal correlation, and environmental noise. We define the state space of $\mathcal{T}(\cdot)$ as $\boldsymbol{\hat s}_i = \left[ m_i, \hat\tau_i, \sigma^2 \right]$, where $\hat\tau_i$ is the idle time of the Tx from the time of last transmission to the current time slot $i$, i.e.,
\begin{equation}
	\hat\tau_{i+1} = \left\{ \begin{array}{ll}
		\hat\tau_{i} + \tau,&{\rm if}\ a_i = 0,\\
		0,&{\rm if}\  a_i = 1.
	\end{array} \right.
\end{equation}
 Next, we design the robotic control gain policy $\mathcal{F}(\cdot)$ to adjust the robotic control gain $K_i$ based on transmission decision $a_i$, noise variance $\sigma^2$, and the status difference $\Delta \boldsymbol{x}_i$, as discussed in \eqref{control_policy}. Therefore, the state space of $\mathcal{F}(\cdot)$ is defined as $\boldsymbol{s}_i = \left[ \Delta \boldsymbol{x}_i, a_i, \sigma^2 \right]$. 
 Accordingly, the state space of MADRL can be expressed as $S = \{\boldsymbol{\hat s}_i, \boldsymbol{s}_i\}$.

 \subsubsection{Action Space}
 
 The outputs of $\mathcal{T}(\cdot)$ and $\mathcal{F}(\cdot)$ are defined as action space, i.e., $a_i \in \{0,1\}$ and $k_i \in \left[ -1, 1 \right] $, where \(k_i\) represents the value after normalization by \(K_i\). The action $a_i$ is discrete and $k_i$ is continuous. Therefore, the action space of MADRL can be expressed as $A = \{a_i, k_i\}$. Moreover, the action of $\mathcal{T}(\cdot)$ is also the state of $\mathcal{F}(\cdot)$.

 \subsubsection{Hybrid Reward}
 We aim to minimize the communication overhead while ensuring precise teleoperation of the robot arm based on the SI design. Therefore, the Huber Loss \cite{huberloss} is adopted to evaluate the control performance of the robot arm, which can be expressed as
 \begin{equation} \label{error}
 	v_i = \left\{ \begin{array}{ll}
 		\frac{1}{2}{e_i^2},&{\rm if}\ e_i < \delta_l , \\
 		\delta_l e_i - \frac{1}{2}{\delta_l ^2},&{\rm if}\ e_i \geq \delta_l ,
 	\end{array} \right.
 \end{equation}
 where $e_i=\Vert \boldsymbol{\bar{x}}_i - \boldsymbol{\hat x}_i \Vert$ is the control error between the actual status of the robot arm and the original status from the Geomagic Touch X device in time slot $i$ (see the sensing subsystem in Fig. \ref{system_diagram}). In addition, $\delta_l$ is a control accuracy parameter. 
 Compared to mean squared error (MSE) and mean absolute error (MAE), Huber Loss demonstrates greater robustness in handling outliers. This is due to its use of a squared and linear penalty for errors below and above the threshold, i.e., $e_i < \delta_l$ and $e_i \geq \delta_l$, respectively. This characteristic enables Huber Loss to achieve combined advantages of MSE and MAE, making it a more balanced loss function in robot control.
 The reward of the information update policy at time slot $i$ is defined as
 \begin{equation} \label{reward}
 	r_i = \left\{ \begin{array}{ll}
 		- v_i,&{\rm if}\ a_i = 1,\\
 		{c_2}({c_1} - v_i),&{\rm if}\  a_i = 0,
 	\end{array} \right.
 \end{equation}
 where $c_1 = \frac{1}{2}{\delta_l^2}$ and $c_2 = \frac{{{\delta _u} - 0.5\delta_l }}{{{\delta _u} - \delta_l }}$ are constant parameters, and $\delta_u > \delta_l$ is associated with the control error upper bound. This reward function has the following properties.
 
 \begin{property}\label{property1}
 	Based on \eqref{error} and \eqref{reward}, we have the following properties:
 	\begin{equation} 
 		r_i\left(a_i=0| e_i < \delta_l\right) > r_i\left(a_i=1| e_i < \delta_l\right),
 	\end{equation}
 	\begin{equation} 
 		r_i\left(a_i=0|e_i=\delta_u\right) = r_i\left(a_i=1|e_i=\delta_u\right),
 	\end{equation}
 	\begin{equation} 
 		r_i\left(a_i=0|e_i>\delta_u\right) < r_i\left(a_i=1|e_i>\delta_u\right).
 	\end{equation}
 \end{property}
 \begin{proof}
 	See Appendix \ref{appb1}.
 \end{proof}

 Based on Property \ref{property1}, it can be shown that the reward for $a_i=0$ surpasses that for $a_i=1$ if $e_i <\delta_l$. 
 This indicates that more rewards can be obtained without activating the communication if $e_i <\delta_l$, thereby reducing the communication overhead without violating the control error requirement.
 In addition, if $e_i \geq \delta_u$, then transmission should be activated to mitigate the control error, thereby enhancing the overall reward. Furthermore, if $\delta_l \leq e_i \leq \delta _u$, there exists a trade-off between minimizing the communication overhead and enhancing the control performance, which necessitates careful optimization to reconcile it. It follows that implementing the reward design in \eqref{reward} can properly characterize the control performance and the communication overhead.
 Therefore, the reward space of MADRL can be expressed as $R = \{r_i, v_i\}$.
 
 \subsubsection{Long-Term Reward}
 
 To solve the problem $\mathcal{P}$, we set the training objective function of DRL as maximizing the long-term reward, which can be expressed as
 \begin{align}
 	Q \left(S, A\right) = \max \mathbb{E} \left[ \sum\limits_{i=0}^{\infty} \gamma^i r_i \mid S_0 = S, A_0 = A\right],
 \end{align}
 where $\gamma$ is the discount factor.

 \subsubsection{Proposed HR-MADRL Architecture}

 \begin{figure}[!t]
 	\centering
 	\vspace{-0.0cm}
 	\setlength{\abovecaptionskip}{-0.0cm}
 	\setlength{\belowcaptionskip}{-0.2cm} 
 	\includegraphics[scale=0.85]{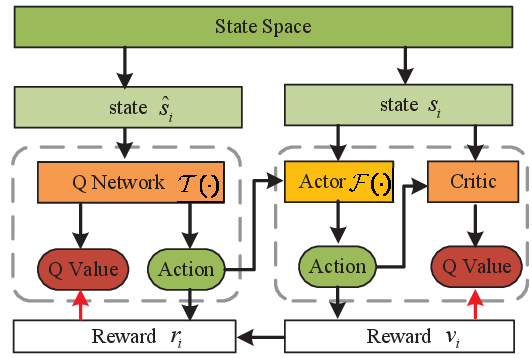}
 	\caption{Proposed HR-MADRL architecture.}
 	\label{fig-HR-MADRL}
 	\vspace{-0.4cm}
 \end{figure}

 As shown in Fig. \ref{fig-HR-MADRL}, we present the network architecture of HR-MADRL. Agent $\mathcal{T}\left(\cdot\right)$ produces discrete action using a  Q-network, while agent $\mathcal{F}\left(\cdot\right)$ generates continuous action utilizing an Actor-Critic (AC) frameworks. The agents operate asynchronously: agent $\mathcal{T}\left(\cdot\right)$ first determines its action; then, the action of $\mathcal{T}\left(\cdot\right)$ is input into the agent $\mathcal{F}\left(\cdot\right)$. Agent $\mathcal{F}\left(\cdot\right)$ interacts with the environment, receives reward $v_i$, and sends feedback to agent $\mathcal{T}\left(\cdot\right)$ to compute its environmental reward $r_i$. These rewards and Q-values are used to update the network parameters via gradient backpropagation.

 To enhance network learning, we employ a twin-network structure in our HR-MADRL. 
 HR-MADRL is an off-policy algorithm with twin-network structure, consisting of two dueling deep Q-networks (DQNs) \cite{dueling_dqn}, and six DNNs.
 As shown in Fig. \ref{fig-madrl}, the target networks mirror the architecture of the current networks, similar to the dual-network approach in the twin delayed deep deterministic policy gradient (TD3) algorithm \cite{td3}.
 The neural parameters are denoted as  ${\boldsymbol{\Theta}}_d$, ${\boldsymbol{\Theta}}_d^{\prime}$, $ {\boldsymbol{\theta}}_{\mu},{\boldsymbol{\theta}}_{q_1}, {\boldsymbol{\theta}}_{q_2}, {\boldsymbol{\theta}}_{\tilde\mu}, {\boldsymbol{\theta}}_{\tilde{q}_1}, {\boldsymbol{\theta}}_{\tilde{q}_2}$, respectively. The actor network and critic (or Q) network are respectively denoted as $\mu(\cdot)$ and $Q(\cdot)$.

 \begin{figure}[!t]
 	\centering
 	\vspace{-0.0cm}
 	\setlength{\abovecaptionskip}{-0.0cm}
 	\setlength{\belowcaptionskip}{-0.2cm} 
 	\includegraphics[scale=0.65]{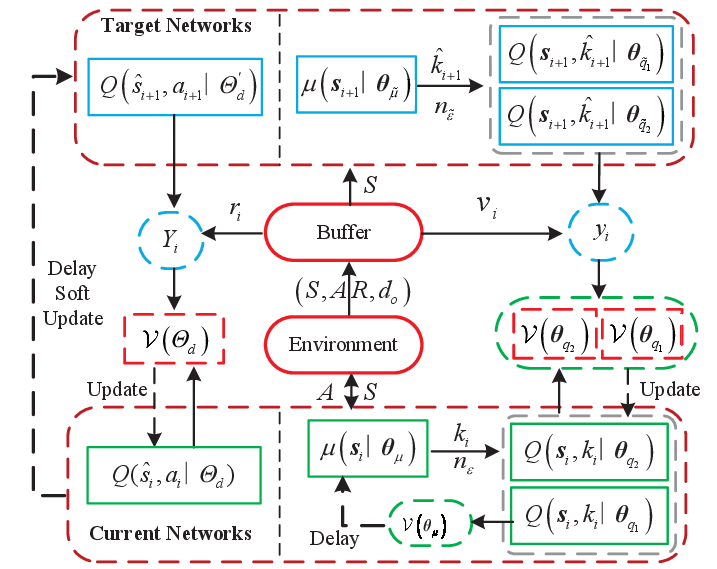}
 	\caption{Proposed HR-MADRL learning structure.}
 	\label{fig-madrl}
 	\vspace{-0.4cm}
 \end{figure}

 \begin{figure}[!t]
 	\centering
 	\vspace{-0.cm}
 	\setlength{\abovecaptionskip}{-0.1cm}
 	\setlength{\belowcaptionskip}{-0.1cm} 
 	\includegraphics[scale=0.75]{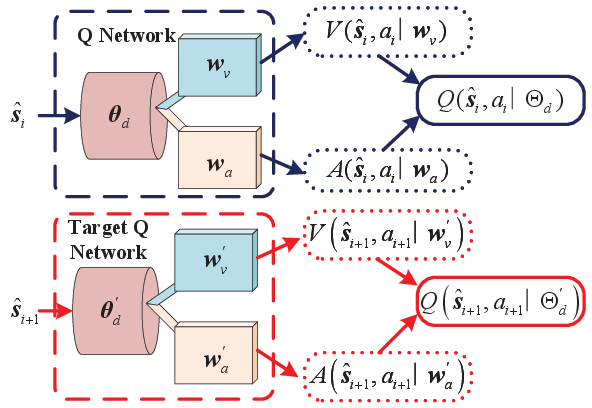}
 	\caption{The Q and target Q network architectures.}
 	\vspace{-0.2cm}
 	\label{fig_d3qn}
 \end{figure}

 As shown in Fig. \ref{fig_d3qn}, the Q network and target Q network with the parameters ${\boldsymbol{\Theta}}_d=\left( {\boldsymbol{\theta}}_d, \boldsymbol{w}_v, \boldsymbol{w}_a \right)$ and ${\boldsymbol{\Theta}}_d^{\prime} = \left( {\boldsymbol{\theta}}_d^{\prime}, \boldsymbol{w}_v^{\prime}, \boldsymbol{w}_a^{\prime} \right) $, respectively. 
 The value function of the Q network (or the target Q network), i.e., $V \left( \boldsymbol{\hat s}_i \mid {\boldsymbol{\theta}}_d, \boldsymbol{w}_v \right)$ (or $V \left( \boldsymbol{\hat s}_i \mid {\boldsymbol{\theta}}_d, \boldsymbol{w}_v^{\prime} \right)$), is characterized by the parameter $\boldsymbol{w}_v$ (or $\boldsymbol{w}_v^{\prime}$), while the advantage function of the Q network (or the target Q network), i.e., $A \left(\boldsymbol{\hat s}_i, a_i \mid  {\boldsymbol{\theta}}_d, \boldsymbol{w}_a \right)$ (or $A \left(\boldsymbol{\hat s}_i, a_i \mid  {\boldsymbol{\theta}}_d, \boldsymbol{w}_a^{\prime} \right)$), is characterized by the parameter $\boldsymbol{w}_a$ (or $\boldsymbol{w}_a^{\prime}$). Note that we utilize similar symbols for the above two networks for convenience. The output of the adopted network yields the state-action value $Q(\boldsymbol{\hat s}_i, a_i \mid {\boldsymbol{\Theta}}_d)$ by jointly considering the value functions of the two networks, which is given by
 \begingroup\makeatletter\def\f@size{9}\check@mathfonts
 \def\maketag@@@#1{\hbox{\m@th\normalsize\normalfont#1}}%
 \begin{equation}
 	\begin{aligned}
 		Q(\boldsymbol{\hat s}_i, a_i\mid {\boldsymbol{\Theta}}_d)  = &  V \left( \boldsymbol{\hat s}_i \mid {\boldsymbol{\theta}}_d, \boldsymbol{w}_v \right) + A \left(\boldsymbol{\hat s}_i, a_i \mid  {\boldsymbol{\theta}}_d, \boldsymbol{w}_a \right)\\
 		&   - \frac{1}{2} \sum_{n_a\in \{0,1\}} A \left(\boldsymbol{\hat s}_i, a_i=n_a \mid  {\boldsymbol{\theta}}_d, \boldsymbol{w}_a \right) .
 	\end{aligned}
 \end{equation}
 \endgroup
 The Q-learning value can be written as
 \begin{equation}
 	Y_i =   r_i + 
 	\gamma (1-d_o) Q \left(\boldsymbol{\hat s}_{i+1}, a_{i+1} \mid \boldsymbol{\Theta}_d^{\prime}\right)	,
 \end{equation}
 where $r_i$ is defined in \eqref{reward} and $a_{i+1} = \mathcal{T} \left(\boldsymbol{\hat s}_{i+1} \right)$. In addition, ``$d_o = 1$'' denotes the end of an epoch. Otherwise, we set ``$d_o = 0$''. The transmission activation decision $\mathcal{T} \left( \cdot \right)$ in \eqref{dqn} for information update is given by
 \begin{equation}
 	\mathcal{T} \left(\boldsymbol{\hat s}_{i} \right) = \mathop{\arg\max}\limits_{a_i} Q\left(\boldsymbol{\hat s}_{i}, a_i \mid {\boldsymbol{\Theta}}_d \right) .
 \end{equation}
 The loss function of the Q network is given by
 \begin{equation}\label{loss_dqn}
 	\mathcal{V} \left( {\boldsymbol{\Theta}}_d \right) = Y_i - Q(\boldsymbol{\hat s}_i, a_i\mid {\boldsymbol{\Theta}}_d).
 \end{equation}


 For the agent $\mathcal{F\left(\cdot\right)}$, the loss functions of the two critic networks can be expressed as
 \begin{equation}\label{loss_q}
 	\setlength{\abovedisplayskip}{3pt}
 	\setlength{\belowdisplayskip}{3pt}
 	\mathcal{V}\left({\boldsymbol{\theta}}_{q_m}\right)=\left(y_i-Q\left(\boldsymbol{s}_{i}, k_{i} \mid {\boldsymbol{\theta}}_{q_m}\right)\right)^2, m=\{1,2\}.
 \end{equation}
 Accordingly, the cumulative reward is represented as
 \begin{equation}\label{y_i}
 	y_i= -v_i +\gamma(1 - d_o) \min _{m=1,2} Q\left(\boldsymbol{s}_{i+1}, \hat{k}_{i+1} \mid {\boldsymbol{\theta}}_{\tilde{q}_m}\right),
 \end{equation}
 where the reward $v_i$ is defined in \eqref{error}. The predicted next action $\hat{k}_{i+1}$ by the target actor network is expressed as
 \begin{equation}
 	\hat{k}_{i+1}= \operatorname{clip}\left(\mu\left(\boldsymbol{s}_{i+1} \mid {\boldsymbol{\theta}}_{\tilde{\mu}}\right) + n_{\tilde\varepsilon},-1,1 \right),
 \end{equation}
 where $n_{\tilde\varepsilon}=\operatorname{clip}\left(\mathcal{N}\left(0, \sigma_\pi^2\right),-\hat{n}_\epsilon, \hat{n}_\epsilon\right)$ is the truncated Gaussian policy noise with the variance $\sigma_\pi^2$, and the operator $\operatorname{clip}\left(x,x_l,x_h\right)$ is given by
 \begin{equation}
 	\operatorname{clip} \left(x,x_l,x_h\right)= \min\left(\max\left(x,x_l\right),x_h\right).
 \end{equation}
 The objective of the actor network is to maximize the evaluation derived from the critic. Therefore, the actor loss is defined as 
 \begin{equation}\label{loss_a}
 	\setlength{\abovedisplayskip}{3pt}
 	\setlength{\belowdisplayskip}{3pt}
 	\mathcal{V}\left({\boldsymbol{\theta}}_\mu\right)=-Q\left(\boldsymbol{s}_{i}, k_{i} \mid {\boldsymbol{\theta}}_{q_1}\right),
 \end{equation}
 where 
 \begin{equation}\label{ki}
 	k_{i} = \operatorname{clip}\left(\mu\left(\boldsymbol{s}_{i} \mid {\boldsymbol{\theta}}_{\mu}\right) + n_\varepsilon, -1,1 \right)
 \end{equation}
 is the output of the actor network with exploration noise $n_\varepsilon\sim \mathcal{N} \left(0,\delta_a^2\right)$.

 By linearly mapping the actor network's output value $\mu\left(\boldsymbol{s}_{i} \mid {\boldsymbol{\theta}}_{\mu}\right)$ from the range of $[-1,1]$ to the domain of the control gain $[0,K_{max}]$, the control policy $\mathcal{F}\left( \cdot \right)$ in \eqref{control_policy} can be obtained as
 \begin{equation}
 	\setlength{\abovedisplayskip}{3pt}
 	\setlength{\belowdisplayskip}{3pt}
 	\mathcal{F}\left( \boldsymbol{s}_{i} \right) = \frac{1}{2} K_{max} (1+\mu\left(\boldsymbol{s}_{i} \mid {\boldsymbol{\theta}}_{\mu}\right)),
 \end{equation}
 where $K_{max}$ is maximum control gain.

\subsection{Overall Training}
Based on the network structures designed previously, we next train them through decentralized MADRL in a virtual environment. The experience replay buffer is employed to store the experience data for HR-MADRL, denoted as $\mathcal{B}$.
To speed up the process of the experiment, we employ the open-source physics engine panda-gym \cite{gallouedec2021pandagym} to replace actual Franka Emika Robot arm for training. Panda-gym comprises a suite of reinforcement learning environments specifically tailored for the Panda and integrated with OpenAI Gym, which can speed up the validation of the proposed method.

At the beginning of the training, the system requires continuous data transmission to the Rx for an initial time duration of $T$. This is crucial since the SFR relies on consecutive inputs with time duration of $T$ to precisely extract semantic features to guarantee accurate prediction. Meanwhile, this time duration allows for a the response preparation phase for controlling the robot arm. 
To stabilize the training outcomes, the exploration noise $n_\varepsilon$ in HR-MADRL gradually decreases as training progresses, with its attenuation parameter denoted as $\eta_a$.
Furthermore, an $\epsilon$-greedy strategy \cite{epsilon-greedy} is utilized to obtain the transmission decision for information update, where $\epsilon$ decays $\epsilon_d$ at the end of each episode in training. 
In the $\epsilon$-greedy strategy, we set the probability of $a_i=0$ to be different from that of $a_i=1$, and set $a_i \sim B(p_a)$ for a certain parameter $p_a$.
This variance helps produce asymmetric observed data in the experience buffer $\mathcal{B}$, thereby amplifying the importance of specific data. As our objective is to significantly reduce the communication overhead while maintaining a good control performance by Q-network, we can configure the probability of $a_i=0$ being greater than that of $a_i=1$, i.e, $p_a<0.5$.

\begin{algorithm}[h]
	\caption {Overall Training Process}
	\label{alg:isc3} 
	\textbf{Preliminary}: $\mathcal{I}(\cdot)$ is obtained by invoking Algorithm \ref{alg:MINE}. \\
	\textbf{Preliminary}: $\mathcal{R}(\cdot)$ is obtained by invoking Algorithm \ref{alg:LSTM}. \\
	\textbf{Initialization}: Initialize the parameters of networks, i.e.,
	${\boldsymbol{\Theta}}_d$, ${\boldsymbol{\theta}}_\mu$, ${\boldsymbol{\theta}}_{q1}$, and  ${\boldsymbol{\theta}}_{q2}$, and those of the target network, i.e.,  ${\boldsymbol{\Theta}}_d^{\prime} \leftarrow {\boldsymbol{\Theta}}_d $, ${\boldsymbol{\theta}}_{\tilde{\mu}} \leftarrow {\boldsymbol{\theta}}_\mu$,  ${\boldsymbol{\theta}}_{\tilde{q}_1} \leftarrow {\boldsymbol{\theta}}_{q_1}$, and ${\boldsymbol{\theta}}_{\tilde{q}_2} \leftarrow {\boldsymbol{\theta}}_{q_2}$. Let $\epsilon \leftarrow 1$.
	\begin{algorithmic}[1]
		\For{$episode=1,2,\cdots,episode_{max}$ } 
		\State Initialize the training environment. Let $\delta_a^2 \leftarrow \eta_a \delta_a^2$ and randomly set the overall AWGN variance $\sigma^2$;  
		\For{$i = 0,1,\cdots$}
		\If{$i > T$}
		\State Generate $b$ according to $b \sim B(\epsilon)$;
		\If{$b > 0$}
		\State Select $a_i$ according to $a_i \sim B(p_a)$;
		\Else
		\State $m_i = \mathcal{I} \left( \boldsymbol{\hat x}_i, \boldsymbol{x}_t \right)$ ,
		$a_i = \mathcal{T} \left( \boldsymbol{\hat s}_{i} \right)$ ;
		\EndIf
		\Else
		\State $a_i = 1$;
		\EndIf
		
		\State \textbf{if} $a_i=1$ \textbf{then} $\boldsymbol{x}_t \leftarrow \boldsymbol{\hat x}_i$ \textbf{end if};
		
		\State Obtain state $S$, action $A$, reward $R$, and flag $d_o$ from the environment;
		
		\State Store $\left(S,A,R,d_o\right)$ in $\mathcal{B}$;
		\EndFor
		
		\For{ $j=1,2,\cdots,N_u$ }
		\State Randomly sample $N$ batches of tuple $\left(S,A,R,S_{\text{next}},d_o\right)$ from $\mathcal{B}$;
		\State Calculate the loss in \eqref{loss_dqn}, \eqref{loss_q} and \eqref{loss_a};
		\State Update ${\boldsymbol{\Theta}}_d$, ${\boldsymbol{\theta}}_{q1}$, and ${\boldsymbol{\theta}}_{q2}$ and  via \eqref{Q_update} and \eqref{q_update};
		\If{ $j\mod\tau_d = 0$ }
		\State Update ${\boldsymbol{\theta}}_{\mu}$, ${\boldsymbol{\Theta}}_d^{\prime}$, ${\boldsymbol{\theta}}_{\tilde{q}_1}$, ${\boldsymbol{\theta}}_{\tilde{q}_2}$, and ${\boldsymbol{\theta}}_{\tilde{\mu}}$ via \eqref{a_update} and \eqref{taget_update};
		\EndIf
		\EndFor
		\State Update $\epsilon$: $\epsilon = \max \{\epsilon-\epsilon_d, \epsilon_{min} \}$
		\EndFor
	\end{algorithmic} 
\end{algorithm}

At the end of each episode, we update the networks of HR-MADRL as follows.
By minimizing \eqref{loss_dqn}, \eqref{loss_q} and \eqref{loss_a} using the gradient descent method, we can update the network parameters as follows:
\begin{equation}\label{Q_update}
	{\boldsymbol{\Theta}}_d \leftarrow {\boldsymbol{\Theta}}_d + \nabla_{{\boldsymbol{\Theta}}_d} \mathcal{V} \left( {\boldsymbol{\Theta}}_d \right) ,
\end{equation}
\begin{equation}\label{q_update}
	{\boldsymbol{\theta}}_{qm} \leftarrow {\boldsymbol{\theta}}_{qm} + \nabla_{{\boldsymbol{\theta}}_{qm}} \mathcal{V} \left( {\boldsymbol{\theta}}_{qm} \right), m=\{1,2\}.
\end{equation}
\begin{equation}\label{a_update}
	{\boldsymbol{\theta}}_{\mu} \leftarrow {\boldsymbol{\theta}}_{\mu} + \nabla_{{\boldsymbol{\theta}}_{\mu}} \mathcal{V} \left( {\boldsymbol{\theta}}_{\mu} \right),
\end{equation}
While the target networks are updated via a soft update policy with a delay $\tau_d$, which can be represented as
\begin{equation}\label{taget_update}
	\begin{aligned}
		{\boldsymbol{\Theta}}_d^{\prime} & \leftarrow \rho {\boldsymbol{\Theta}}_d^{\prime} + (1-\rho) {\boldsymbol{\Theta}}_d , \\
		{\boldsymbol{\theta}}_{\tilde{q}_m} & \leftarrow \rho {\boldsymbol{\theta}}_{q_m}+(1-\rho) {\boldsymbol{\theta}}_{\tilde{q}_m}, m=\{1,2\} , \\
		{\boldsymbol{\theta}}_{\tilde{\mu}} & \leftarrow \rho {\boldsymbol{\theta}}_\mu+(1-\rho) {\boldsymbol{\theta}}_{\tilde{\mu}} ,\\
	\end{aligned}
\end{equation}
where $\rho \ll 1$ is a discount factor.

The details of the overall training algorithm are outlined in Algorithm \ref{alg:isc3}. 
It is worth noting that the HR-MADRL undergo $N_u$ training iterations after each episode completion, as shown from line 18 to line 25 in Algorithm \ref{alg:isc3}, where the actor network and the target networks employ a delayed update policy with a delay parameter $\tau_d$, as shown from line 22 to line 24.

Algorithm \ref{alg:isc3} effectively addresses the problem $\mathcal{P}$. By pre-training the models \( \mathcal{I} \) and \( \mathcal{R} \), we ensure that the decision-making processes of agents \( \mathcal{T} \) and \( \mathcal{F} \) adhere to the Markov property. Specifically, at the Tx, the decision state input for agent \( \mathcal{T} \) is modified from \( \hat{\boldsymbol{x}}_i \) to \( m_i \), which represents the semantic feature differences between past and current states. At the Rx, we employ SFR to reconstruct semantic features over the temporal dimension, thereby directly mapping the influence of past states onto the current state. This approach enables agent \( \mathcal{F} \) to make decisions based on the current predicted state. Furthermore, the innovative hybrid reward function theoretically ensures an accurate evaluation of the decision quality of agents \( \mathcal{T} \) and \( \mathcal{F} \), thereby guaranteeing the convergence of the training process. Through these measures, we can efficiently solve problem $\mathcal{P}$ while ensuring compliance with control error constraints.


\section{Experimental Results}
In this section, we carry out on-site experiments to evaluate the efficacy of our proposed method as compared to other baseline schemes.

\subsection{Experiment Setting}

\subsubsection{Dataset}

\begin{figure}[t]
	\centering
	\includegraphics[trim=0.15cm 0.0cm 0.3cm 0.3cm,clip,scale=0.75]{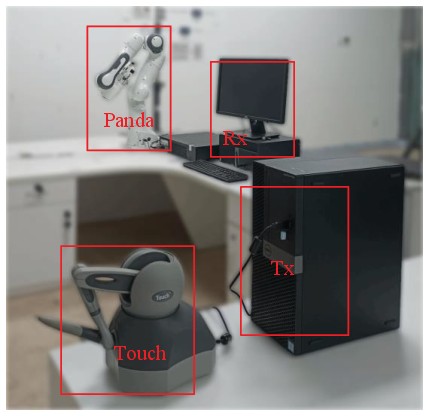}
	\caption{A teleoperation system. }
	\vspace{-0.2cm}
	\label{fig:photo}
\end{figure}

\addtolength{\footskip}{0.02cm}
In the experiment, the adopted data was gathered from a teleoperation system as shown in Fig. \ref{fig:photo}. Specifically, a tactile hardware device (i.e., Geomagic Touch X) serves as the master device in the teleoperation, and the position of its Touch nib is collected by a computer in real-time and treated as the sensed data. Then, a transmission decision is generated by SI in the SFE computing module at the Tx. If the transmission is activated, the corresponding data is sent to the Rx via wireless local area network (WLAN) for the control of the slave device, i.e., the Franka Emika Robot Arm named {\it Panda}. Remote control is established by mapping the three-dimensional position of the Touch's end point to that of the robotic arm's end point. 
However, the spatial scale of the Touch's motion may not align with that of Panda. Hence, we introduce the following linear mapping of Touch's space into Panda's space,
\begingroup\makeatletter\def\f@size{9}\check@mathfonts
\def\maketag@@@#1{\hbox{\m@th\normalsize\normalfont#1}}%
\begin{equation} \label{touch2panda}
	\boldsymbol{\hat x}_{b,i}(j) = \boldsymbol{x}_{b,l}(j) + \frac{\boldsymbol{x}_{b,u}(j)-\boldsymbol{x}_{b,l}(j)}{\boldsymbol{x}_{t,u}(j)-\boldsymbol{x}_{t,l}(j)} \left( \boldsymbol{\hat x}_i(j) -\boldsymbol{x}_{t,l}(j)\right) ,
\end{equation}
\endgroup
where $\boldsymbol{\hat x}_{b,i}$ is the location that the Touch space maps to the space of the robotic arm, and $j=\{1,2,3\}$ represent the $x$, $y$, and $z$ axes in three dimensions, respectively.
 In addition, we set $\boldsymbol{x}_{b,l} = [-200,-200,0]$, $\boldsymbol{x}_{b,u} = [200,200,400]$, $\boldsymbol{x}_{t,l} = [-105,-100,-220]$, $\boldsymbol{x}_{t,u} = [263,150,100]$ (all in millimeter (mm)). Furthermore, we adopt the peak status-to-noise-ratio (PSNR) to indicate the strength of noise or interference, where the quantification method in \cite{changbo_jsac} is used to convert the physical coordinate into information domain. In this regard, PSNR means the ratio of the status scale to the environmental noise variance, i.e.,
\begin{equation}
	\text{PSNR} = 10 \lg \frac{\Vert \boldsymbol{x}_{b,u}-\boldsymbol{x}_{b,l} \Vert^2}{\sigma^2}  \left(\text{dB}\right) .
\end{equation}
We assume that the status scale is a constant. As such, a smaller PSNR should imply stronger noise and interference.

The sensing frequency for the status update at the Touch is 120 Hz, where the average transmission rate to robot arm via WLAN is 21 Hz. A total of 820,000 samples are collected in about 10.8 hours, where 700,000 samples are allocated for the training set and the remaining 120,000 samples for the test set.

%

\subsubsection{Metrics}

In this experiment, we focus on the communication overhead and the error of remote control for the teleoperation. The communication overhead is measured by the DC $\eta_{dc}$ in (\ref{equ2}).
The error of remote control is defined as
\begin{equation}
	S_e = \mathbb{E}\left[\Vert \boldsymbol{\hat x}_{b,i} - \boldsymbol{\bar{x}}_i \Vert \right]   ,
\end{equation}
where $\boldsymbol{\hat x}_{b,i}$ is derived from the mapping of $\boldsymbol{\hat x}_i$ in \eqref{touch2panda}.

\subsubsection{Networks Architecture}
The parameters of the networks are listed in Table \ref{network_para}. The MINE, Q networks, and AC networks utilize a fully-connected (FC) layer with the ReLU activation function. Meanwhile, the feature encoder-decoder computing unit employs an LSTM network architecture with 3-state inputs and 64 hidden layers. Since the coordinations cannot represent all motion characteristics, velocity in the teleoperation is adopted as the input in the MINE, and LSTM networks, which is computed by the distance of two consecutive statuses and denoted as $\boldsymbol{V}_i = \boldsymbol{\hat x}_i - \boldsymbol{\hat x}_{i-1}$. 

\begin{table}
	\caption{Parameters of The Learning Networks}
	\centering
	\begin{tabular}{|l|l|l|}
		\hline
		& Layer & Dimensions \\
		\hline
		MINE & FC$\times$5 & $(3,64,128,128,64,1)$  \\
		Encoder & LSTM $\times$3 & input=3,hidden=64  \\
		Decoder & LSTM $\times$3 & input=3,hidden=64  \\
		Q Network &  FC$\times$5 & $(3,64,128,128,64, 2)$ \\
		Actor & FC$\times$5 & $(5,64,128,128,64,1)$ \\
		Critic & FC$\times$5 & $(6,64,128,128,64,1)$ \\
		\hline
	\end{tabular}\label{network_para}
\end{table}

\subsubsection{Initial Parameter Setting}

The primary experimental parameters are listed in Table \ref{para}. An Adam optimizer with a learning rate 0.001 is adopted in the learning networks. In addition, an experience replay strategy is employed during the training procedures for HR-MADRL, where an experience buffer size of $\mathcal{B} = 1,000,000$ is used in this strategy. The batch size $N$ is set to 1024.

\begin{table}
	\caption{Parameters of Experiment}
	\centering
	\begin{tabular}{|l|l|l|l|l|l|}
		\hline
		$\gamma$ & $0.99$ & $\rho$ & $0.005$ & $\epsilon_d$ & $7\times10^{-4}$ \\
		$\epsilon_{min}$ & $0.05$ & $\delta_a^2$ & $0.2$ & $\delta_\pi^2$ & $0.2$  \\
		$\hat{n}_\epsilon$ & $0.1$ & $K_{max}$ & $30$ & $T$ & $80$\\
		$M$ & $20$ & $\eta_a$ & $0.999$ & $\tau_d$ & $3$\\
		$N_u$ & $20$ & $p_a$ & $0.4$ & & \\
		\hline
	\end{tabular}\label{para}
\end{table}

\subsection{Performance Evaluation}

In this subsection, we show the performance of our proposed method versus several benchmark schemes with independent communication and control designs, i.e.,
\begingroup
\allowdisplaybreaks
\begin{itemize}
  \item Case 1: The control information is transmitted once generated by the Touch device.
  \item Case 2: The control information of the Touch is transmitted every $S$ samples. 
  
  \item Case 3: The control information is transmitted once generated by the Touch device, but the computing unit $\mathcal{F}(\cdot)$ trained by TD3 \cite{td3} for Panda's control process is added to dynamically adjust the control gain at the Rx.
   
  \item Case 4: The LSTM-based encoder-decoder computing unit for SI reconstruction at the Rx is not considered in our proposed method. 
  
  \item Case 5: Computing unit $\mathcal{F}(\cdot)$ for the robot control process optimization is not considered in our proposed method, and the computing unit $\mathcal{T}(\cdot)$ is trained by Double Dueling Deep Q-Network (D3QN) \cite{double_dqn,dueling_dqn}.
  
  \item Case 6: The proposed HR-MADRL architecture is similarly employed. However, in contrast to the proposed method, the computing unit $\mathcal{F}(\cdot)$ of the robot is trained utilizing a non-deterministic policy algorithm, the Soft Actor-Critic (SAC) approach \cite{sac}.

  \item Case 7: The proposed HR-MADRL architecture is similarly employed. However, unlike the proposed method, the computing units $\mathcal{T}(\cdot)$ and $\mathcal{F}(\cdot)$ are trained utilizing an on-policy algorithm, the Proximal Policy Optimization (PPO) approach \cite{ppo}.
  
\end{itemize}
\endgroup

\begin{figure}[!t]
	\centering
	\vspace{-0.cm}
	\setlength{\abovecaptionskip}{-0.1cm}
	\setlength{\belowcaptionskip}{-0.1cm} 
	\includegraphics[trim=0.15cm 0.0cm 0.3cm 0.3cm,clip,scale=0.75]{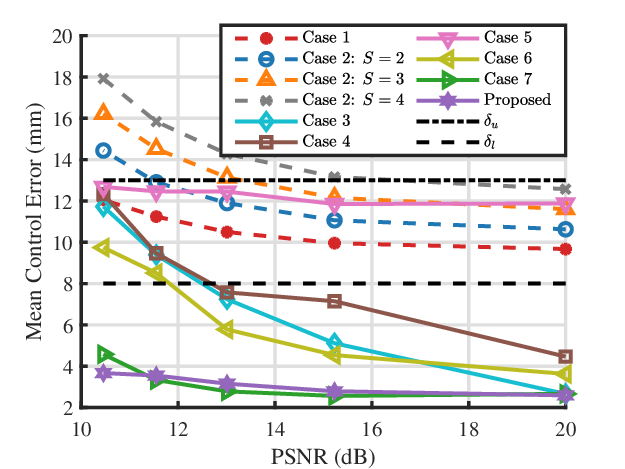}
	\caption{Robot control error versus PSNR. }
	\vspace{-0.1cm}
	\label{fig:error}
\end{figure}

\begin{figure}[!t]
	\centering
	\vspace{-0.cm}
	\setlength{\abovecaptionskip}{-0.1cm}
	\setlength{\belowcaptionskip}{-0.2cm} 
	\includegraphics[trim=0.05cm 0.0cm 0.3cm 0.3cm,clip,scale=0.75]{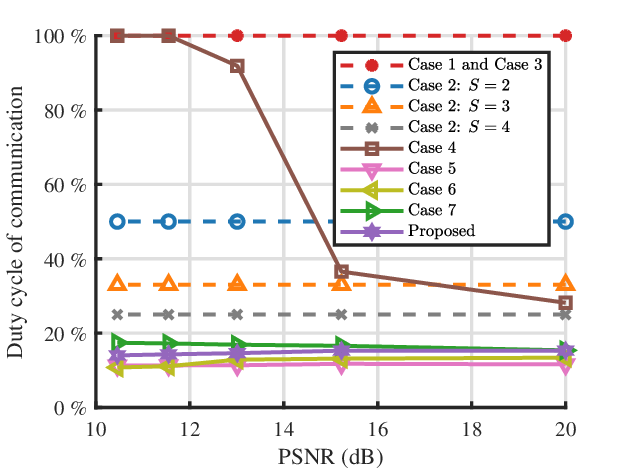}
	\caption{Communication DC versus PSNR. }
	\vspace{-0.2cm}
	\label{fig:rate}
\end{figure}

Fig. {\ref{fig:error} shows the mean control error by the proposed method and the above benchmarks. In Case 2, we consider sampling interval $S=2$, $3$, and $4$. It is observed that the mean control errors by all schemes decrease monotonically with PSNR. This is because a larger PSNR implies smaller disturbance at the Panda, allowing for its more precise control. 
Moreover, it also implies lower electromagnetic noise on the status update information, which helps reduce the control error as well.
Hence, a larger PSNR helps reduce the mean control error in both physical and information domains.

It is also observed from Fig. {\ref{fig:error} that the control error in Case 2 is larger than that in Case 1. In addition, the control error becomes larger with increasing $S$, since a smaller information update period leads to more timely control information update and thus enables better control performance. Moreover, compared with other benchmarks, our proposed method yields the best control performance with the smallest control error. Nonetheless, Case 3 is observed to converge to the proposed method when the PSNR is sufficiently large, which implies that the added control process optimization can achieve  optimal control performance in this regime, at the cost of highly frequent information transmission from the Tx to the Rx.

Fig. \ref{fig:rate} shows the communication DC versus the PSNR. It is noted that the duty cycles of both Case 1 and Case 3 is $100\%$ since the transmission is always activated for every sample information generated at the Touch. In Case 2, the DC is observed to decrease with $S$, since a larger $S$ results in smaller transmission activation. However, the control error increases accordingly.
Furthermore, Case 4 is observed to decrease dramatically when the PSNR is larger than $13$ dB. This implies that there exists a threshold for the PSNR, below which the received information is difficult to be reconstructed in the information domain, and the control performance is only dominated by the control algorithm, i.e., the computing unit $\mathcal{F}(\cdot)$. In contrast, if the PSNR is sufficiently high, the received information can be used for the motion control, and the control performance is determined by both the computing unit $\mathcal{F}(\cdot)$ and the received information from the Tx.

It is also observed from Fig. \ref{fig:rate} that, Case 5 yields the smallest communication duty among all schemes considered. This is reasonable since the requirement of control performance without the computing unit $\mathcal{F}(\cdot)$ is lower; thus, the information update frequency can be reduced without compromising the required control performance. The proposed method is observed to achieve the communication duty of $15\%$, similar to Case 5. It follows that our proposed method can properly balance the control error and the communication overhead.

Fig. \ref{fig:loss} demonstrates that within the HR-MADRL framework proposed in this paper, the three methods can achieve satisfactory convergence performance. Specifically, Case 6 exhibits inferior convergence rates and ultimate performance as compared to Case 7 and the proposed method. Case 7, which utilizes an on-policy strategy for training, displays the fastest convergence speed; however, its inability to utilize historical sample data renders it vulnerable to the current specific samples, resulting in instability during the training process. In contrast, the proposed method attains superior performance in terms of system rewards and control Huber loss. As illustrated in Fig. \ref{fig:rate}, it is evident that under the HR-MADRL framework, Cases 6 and 7, as well as the proposed method are all effective to minimize the communication overhead. Furthermore, Fig. \ref{fig:error} indicates that, in comparison to the proposed method, Case 6 results in larger control errors, while Case 7 yields worse performance compared to the proposed method in the low PSNR regime.

\begin{figure}[!t]
	\centering
	\vspace{-0.cm}
	\setlength{\abovecaptionskip}{-0.1cm}
	\setlength{\belowcaptionskip}{-0.1cm} 
	\includegraphics[trim=0.15cm 0.0cm 0.3cm 0.3cm,clip,scale=0.55]{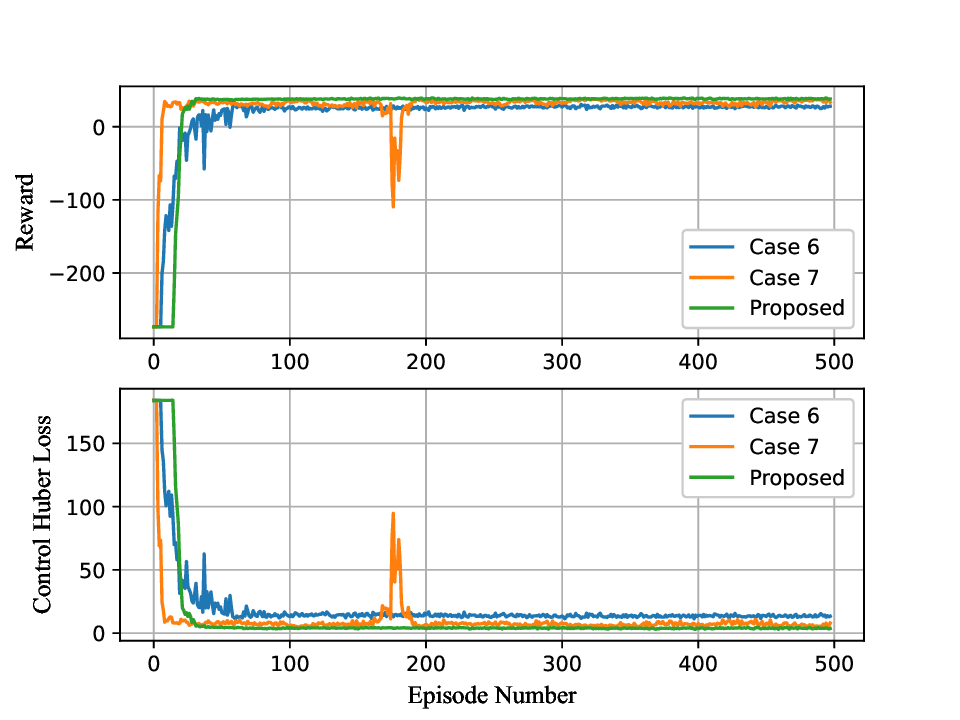}
	\caption{Performance evaluation of Cases 6 and 7 and proposed method in training. }
	\vspace{-0.1cm}
	\label{fig:loss}
\end{figure}

\begin{figure}[!t]
	\centering
	\vspace{-0.cm}
	\setlength{\abovecaptionskip}{-0.1cm}
	\setlength{\belowcaptionskip}{-0.1cm} 
	\includegraphics[trim=0.15cm 0.cm 0.0cm 0.1cm,clip, scale=0.75]{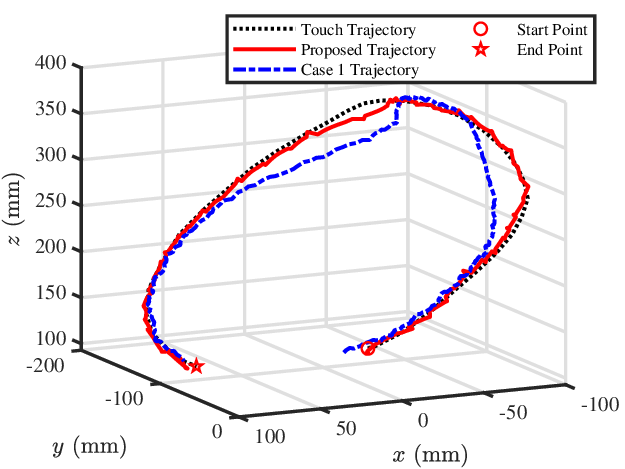}
	\caption{Robot arm's tracking error along its trajectory. }
	\vspace{-0.4cm}
	\label{fig:trajectory}
\end{figure}

Finally, Fig. \ref{fig:trajectory} shows the robot arm tracking error along its trajectory, when the PSNR is 20 dB. It is observed that the robot arm's trajectory under the proposed method is almost the same as that of the Touch (i.e., the sensor at the Tx), while there exist significant deviations between the arm's trajectory under Case 1 and that of the Touch. 
Moreover, the proposed method can reduce the DC of Case 1 by 85\%, as shown in Fig. \ref{fig:rate}. This indicates that our proposed method can achieve a higher motion control accuracy and lower communication overhead at the same time as compared to Case 1.

\subsection{Influence of Hyperparameters}

As illustrated in Figs. \ref{fig:TL_error} and \ref{fig:TL_rate}, we conduct a comparative analysis of the impact of the hyperparameters \(T\) and \(\delta_l\) on the simulation results. It is observed from the figures that as \(T\) increases, both the control error of the robot and the communication DC exhibit a downward trend. This phenomenon can be attributed to the fact that during the training of SFR, a larger value of \(T\) enables the model to learn the characteristics of trajectory changes from a longer history of trajectory data, thereby enhancing the accuracy of prediction. In the experimental results, this is manifested as a lower control error and a lower communication DC for models trained with a larger \(T\) value.
Moreover, it can also be observed from Figs. \ref{fig:TL_error} and \ref{fig:TL_rate} that when \(\delta_l\) is smaller, the control error of the robot decreases, while the communication DC increases. This is because a smaller \(\delta_l\) implies a lower threshold for control error that allows communications to remain inactive, which in turn imposes stricter requirements on control performance. To meet this requirement, the system has to allocate more communication resources, resulting in an increase in the communication DC.

\begin{figure}[!t]
	\centering
	\vspace{-0.cm}
	\setlength{\abovecaptionskip}{-0.1cm}
	\setlength{\belowcaptionskip}{-0.1cm} 
	\includegraphics[trim=0.15cm 0.0cm 0.3cm 0.3cm,clip,scale=0.75]{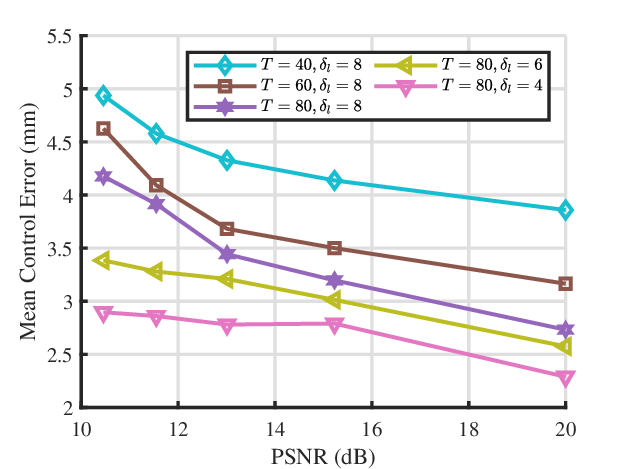}
	\caption{Robot control error versus PSNR. }
	\vspace{-0.1cm}
	\label{fig:TL_error}
\end{figure}

\begin{figure}[!t]
	\centering
	\vspace{-0.cm}
	\setlength{\abovecaptionskip}{-0.1cm}
	\setlength{\belowcaptionskip}{-0.2cm} 
	\includegraphics[trim=0.05cm 0.0cm 0.3cm 0.3cm,clip,scale=0.75]{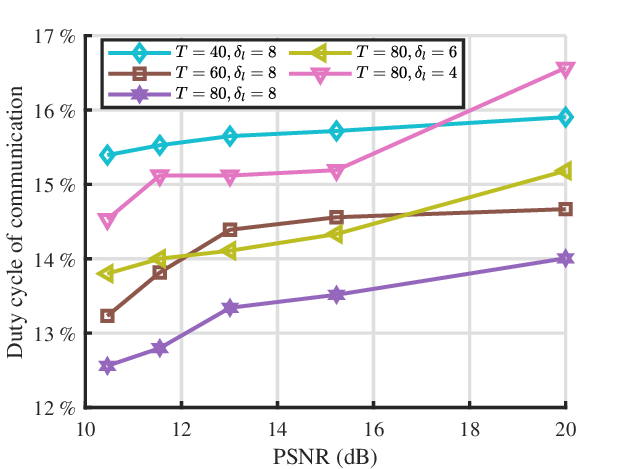}
	\caption{Communication DC versus PSNR. }
	\vspace{-0.2cm}
	\label{fig:TL_rate}
\end{figure}

\subsection{Performance Comparison}

We compare the proposed scheme with two benchmark schemes, including a traditional wireless communication scheme \cite{meng_zhen} and a semantic communication scheme \cite{girgis2024semantic}. The proposed scheme in \cite{meng_zhen} reduces communication overhead and ensures an acceptable model reconstruction error in the metaverse through a sampling, communication, and prediction co-design (SCPC). The proposed scheme in \cite{girgis2024semantic} predicts the controller's state under constrained communication conditions by employing a semantic communication and control co-design (SemCCC). 
Fig. \ref{fig:compare} presents a comparison between the proposed scheme and these two benchmark schemes, demonstrating that our proposed scheme outperforms the two comparison schemes. Specifically, SCPC reduces communication overhead by adjusting the dynamic sampling rate; however, it fails to accurately discern semantic differences between pieces of information, resulting in unnecessary communication transmission. Furthermore, the controller in SCPC is not jointly optimized, which hinders improvements in control accuracy. SemCCC enhances state prediction by extracting semantic features from states, but for extremely concise control information (e.g., control information transmission considered in this paper), extracting semantic features at each sampling does not significantly reduce communication overhead. In summary, our scheme significantly outperforms existing related schemes in terms of both communication efficiency and control accuracy.

\begin{figure}[!t]
	\centering
	\vspace{-0.cm}
	\setlength{\abovecaptionskip}{-0.1cm}
	\setlength{\belowcaptionskip}{-0.2cm} 
	\includegraphics[trim=0.05cm 0.0cm 0.0cm 0.2cm,clip,scale=0.65]{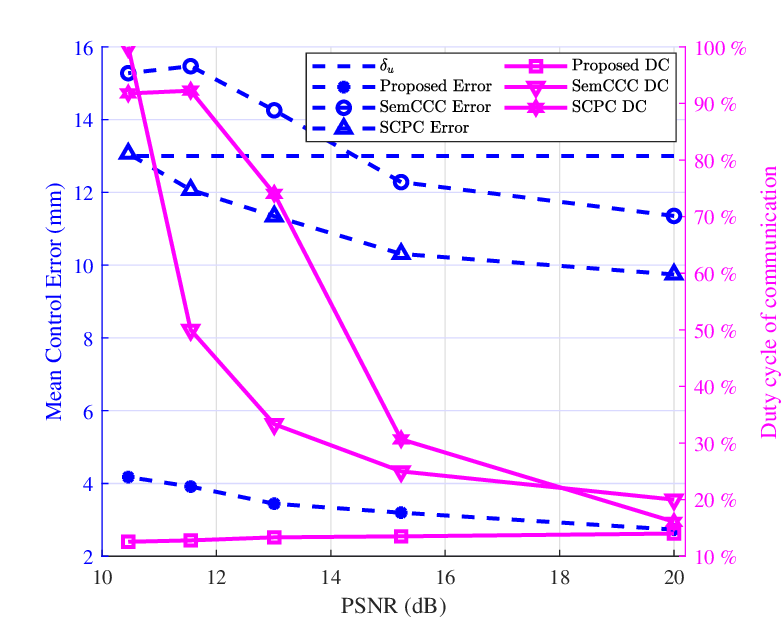}
	\caption{Robot control error and communication DC versus PSNR. }
	\vspace{-0.2cm}
	\label{fig:compare}
\end{figure}

\section{Conclusion and Future Work}
This paper proposed a time-sequence-based SC paradigm by formuating a novel ISC3 architecture to employ SC for task-oriented WCSs. In the architecture, we utilized the MI between the current and previous control information as a criterion to determine the information update policy at the Tx, thereby reducing the communication overhead. Moreover, the SFR was employed at the Rx to predict the control command in the absence of control information transmission from the Tx. In addition, we developed a control gain policy for the controlled target to dynamically adjust the control gain based on the reliability of the control instructions to minimize control errors. We optimized these policies by designing their associated network structures and training process by combining MINE, LSTM and HR-MADRL. Experimental results demonstrated that the robot arm's trajectory by the proposed method was almost the same as that of the sensor at Tx. Moreover, the proposed method can better balance the remote control accuracy and the communication overhead as compared to other benchmark schemes, thus providing a promising solution for future ultra-reliable and low-latency WCSs.

This paper provides a comprehensive exploration of the proposed method for specific wireless teleoperation scenarios. As a pioneering effort in this field, we focus on the ISC3 architecture based on time-sequence SC. The implementation of physical layer design, e.g., the integrated signal processing of communication and sensing, would be left as a future work. 
Another potential future work is introducing novel techniques, e.g., higher carrier frequency to achieve more reliable communications with higher transmitting rate and higher sensing resolution to further improve the overall system performance.
Furthermore, the applications of the proposed method can be extended far beyond the contexts, which can be used in massive scenarios with wireless control in loop, e.g., unmanned aerial vehicles (UAV) for low-altitude economy, remote driving of logistics vehicles, and AR/VR operations in metaverse.


\appendices

\section{Proof of Property 1}\label{appb1}

Based on \eqref{error} and \eqref{reward}, if $e_i < \delta_l$, the reward becomes
	\begin{equation} \label{reward_app}
		r_i = \left\{ \begin{array}{ll}
			- \frac{1}{2}{e_i^2} \leq 0,  &{\rm if}\ a_i = 1,\\
			\frac{{{\delta _u} - 0.5\delta_l }}{{{\delta _u} - \delta_l }} \left( \frac{1}{2}{\delta_l^2} - \frac{1}{2}{e_i^2} 
			\right) >0 ,  &{\rm if}\  a_i = 0.
		\end{array} \right.
	\end{equation}
	Thus, we have $r_i\left(a_i=0| e_i < \delta_l\right) > r_i\left(a_i=1| e_i < \delta_l\right)$.

	If $e_i \geq \delta_l$, the reward can be expressed as
	\begin{equation} \label{reward1_app}
		r_i = \left\{ \begin{array}{ll}
			\frac{1}{2}{\delta_l ^2} - \delta_l e_i  ,&{\rm if}\ a_i = 1,\\
			\frac{{{\delta _u} - 0.5\delta_l }}{{{\delta _u} - \delta_l }} \left( {\delta_l^2} - \delta_l e_i 
			\right) ,&{\rm if}\  a_i = 0.
		\end{array} \right.
	\end{equation}	
	As
	\begin{equation}
		\frac{\partial r_i(a_i=0|e_i\geq \delta_l)}{\partial e_i} = \frac{{0.5\delta_l^2 - {\delta _u}}}{{{\delta _u} - \delta_l }}
	\end{equation}	
	and
	\begin{equation}
		\frac{\partial r_i(a_i=1|e_i\geq \delta_l)}{\partial e_i} = - \delta_l ,
	\end{equation}	
	we have 
	\begin{equation}
		\frac{\partial r_i(a_i=0|e_i\geq \delta_l)}{\partial e_i} < \frac{\partial r_i(a_i=1|e_i\geq \delta_l)}{\partial e_i} 
	\end{equation}	
	Since $r_i\left(a_i=0|e_i=\delta_u\right) = r_i\left(a_i=1|e_i=\delta_u\right)$, it should hold that $r_i\left(a_i=0|e_i>\delta_u\right) < r_i\left(a_i=1|e_i>\delta_u\right)$.

Thus, Property \ref{property1} is proved.

\bibliographystyle{IEEEtran}
\bibliography{IEEEabrv,reference_abrv}

\end{document}